\documentclass[12pt]{article}
\usepackage{authblk}
\usepackage[T1]{fontenc}
\usepackage{tgpagella}
\usepackage{amsmath,amssymb,mathtools,amsthm}
\usepackage{graphicx,epsf,color,xcolor}
\usepackage{enumerate}
\usepackage{natbib}
\usepackage{forest}
\usepackage{float}
\usepackage{url} 
\usepackage{times}
\usepackage[ruled,noend]{algorithm2e}
\usepackage{graphics}
\usepackage{booktabs}
\usepackage{xr}
\usepackage[colorlinks=true, linkcolor=blue, citecolor=blue, filecolor=blue]{hyperref}
\usepackage{setspace}
\usepackage{caption}
\usepackage{subcaption}

\usepackage[margin=1in]{geometry}

\usepackage{times}
\usepackage{lineno,hyperref}

\usepackage{natbib}
\usepackage{caption}
\usepackage{algpseudocode}
\usepackage{soul}
\newtheorem{theorem}{Theorem}[section]

\theoremstyle{remark}

\usepackage{bm}

\modulolinenumbers[5]
\usepackage{enumitem}
\usepackage{booktabs}
\usepackage{multirow}

\def\A{{\rm A}}
\def\B{{\rm B}}

\def\AB{{\rm AB}}
\def\poi{\text{Poi}}

\def\var{\text{Var}}
\def\sc{\mathcal}
\def\bb{\mathbb}

\newif\ifanon
\anonfalse

\begin{document}

\def\spacingset#1{\renewcommand{\baselinestretch}%
{#1}\small\normalsize}
\spacingset{1}

\title{Stochastic Block Covariance Matrix Estimation}
\author[1]{Yunran Chen}
\author[1,4]{Surya T Tokdar}
\author[2,3,4,5]{Jennifer M Groh}
\affil[1]{Department of Statistical Science, Duke University, USA}
\affil[2]{Department of Psychology \& Neuroscience, Duke University, USA}
\affil[3]{Department of Biomedical Engineering, Duke University, USA}
\affil[4]{Duke Institute of Brain Sciences, Duke University, USA}
\affil[5]{Center for Cognitive Neuroscience, Duke University, USA}

\maketitle

\begin{abstract}
Motivated by a neuroscience application we study the problem of statistical estimation of a high-dimensional covariance matrix with a block structure. The block model embeds a structural assumption:  the population of items (neurons) can be divided into latent sub-populations with shared associative covariation within blocks and shared associative or dis-associative covariation across blocks. Unlike the block diagonal assumption, our block structure incorporates positive or negative pairwise correlation between blocks. In addition to offering reasonable modeling choices in neuroscience and economics, the block covariance matrix assumption is interesting purely from the perspective of statistical estimation theory: (a) it offers in-built dimension reduction and (b) it resembles a regularized factor model without the need of choosing the number of factors. We discuss a hierarchical Bayesian estimation method to simultaneously recover the latent blocks and estimate the overall covariance matrix. We show with numerical experiments that a hierarchical structure and a shrinkage prior are essential to accurate recovery when several blocks are present.
\end{abstract}


\newpage

\section{Introduction}

It is well understood that an accurate estimation of a covariance matrix is infeasible without appropriate regularization when the number of variables $p$ is comparable to or exceeds the sample size $n$. Regularization is typically enforced either through assumptions of sparsity, with known or unknown sparsity structures, or by the technique of shrinkage customized to specific composite loss functions and invariance constraints; 
see, for example, \citet{wu2003nonparametric, ledoit2004well, ledoit2020analytical, zou2006sparse, bickel2008regularized,  rajaratnam2008flexible, cai2010optimal, bhattacharya2011sparse}, and the references therein. \citet{cai2016estimating, fan2016overview, pourahmadi2013high} offer comprehensive reviews on the topic.

An approach that combines sparsity with shrinkage, but has received scant attention in the literature, is covariance estimation under stochastic block models. These models posit that the $p$ variables are clustered into $k$ distinct groups (blocks) such that the covariance (or correlation) between any pair of variables is entirely determined by their group memberships. Similar stochastic blockmodels are a popular choice for analyzing network data \citep[see][for a review]{lee2019review}, but it remains underappreciated that such models could bring about unexpected benefits in covariance estimation. \cite{liechty2004bayesian} seem to be the first to use these models to group companies by stock price correlation and group plant phenotypes by size correlation. Crucially, in the second application, they identify two groups of phenotypes with negative cross-block correlation and conclude that plants either emphasize leaf growth or plant growth, but not both. 

Relative to other blocking strategies \citep{perrot2022estimation}, the stochastic block covariance model is unique in its twofold ability to identify suitable clusters and detect cross-block negative correlation -- two features of immense importance to many areas of science. Our study of these models is motivated by neuroscience research where we investigate how a population of neurons coordinate with one another to preserve information from multiple stimuli presented simultaneously. \citet{jun2022coordinated} observe a bimodal distribution of pairwise spike count correlations under dual-stimuli exposures and suggest the existence of subpopulations with synchronous encoding within groups and anti-synchronous activities across -- a conjecture which can be more comprehensively tested with the stochastic block covariance model. Correlation clustering could also be useful beyond hypothesis testing. In finance, for example, accurate clustering of assets could facilitate portfolio diversification and risk mitigation \citep{creal2024bayesian}.

Beyond its scientific appeal, stochastic block covariance estimation is interesting purely as a statistical tool because of its ability to synthesize shrinkage with sparsity. Sparsity is explicitly encoded by the structural assumption of $k$ blocks. One only needs to estimate $k(k+3)/2$ distinct parameters -- $k$ block level variances and $k(k+1)/2$ within and cross-block covariances -- instead of $p(p+1)/2$ free parameters in a $p\times p$ covariance matrix. The role of shrinkage is more implicit. Theorem \ref{theorem1} shows that given the grouping structure, the maximum likelihood estimate of the covariance matrix is obtained by simply averaging the raw sample variances and covariances within and across the groups. This averaging is a shrinkage operation and for it to work one must shrink adaptively by identifying a good partition of the variables so that within and across block averaging does not degrade fidelity between data and the estimate. How to shrink adaptively is the main question we address in this paper (Section \ref{sec:shrink}). Stochastic block covariance models also have close connection with regularized factor models \citep{taylor2017joint, tong2023characterizing} but, arguably, offer a more efficient estimation framework as examined in Section \ref{sec:cfm}.

Clearly, the primary challenge with stochastic block covariance matrix estimation is identification of the blocks which itself constitutes of two interrelated questions. How many blocks are needed? How to assign variables to blocks? It is appealing to address these challenges with a Bayesian approach where one can specify a prior distribution on the space of partitions of the variable indices. But a Bayesian approach presents a new technical challenge. One also needs a prior distribution on the space of block covariance matrices which is a highly constrained subspace of the space of $p\times p$ positive semidefinite matrices. \cite{liechty2004bayesian} use truncation to circumvent the technical issue of constraints but the resulting posterior computation is both complex and limited in scope. Indeed, \cite{liechty2004bayesian} do not fully address the question of {\it how many blocks are needed}; they assume a number is known from the context of the application.

A more fruitful solution is offered by \cite{creal2024bayesian} who exploit a novel matrix factorization result of \cite{archakov2024canonical} to specify a conjugate prior on the block covariance matrix given a known partition of the variables. This specification enables to calculate -- in closed form -- the marginal likelihood score of a given partition. The marginal likelihood is then combined with a partition prior to make inference on the blocks with straightforward Markov chain Monte Carlo computation \citep[e.g.][]{miller2018mixture}. The posterior mean -- which averages over likely partitions -- is the Bayes estimate of the covariance matrix under squared Frobenius loss. 

While such Bayes estimates are a type of shrinkage estimates, they may fail to offer adaptive shrinkage in the learning of the blocks. Conditionally on a given partition, the mean of the conjugate prior provides a focal point of shrinkage of the pairwise covariance values and the prior variance controls the degree of shrinkage. For example one could specify a hyper-inverse Wishart $g$-prior \citep{carvalho2009objective} to shrink towards the conditional maximum likelihood estimate but such a specification would bias the method toward identifying a large number of blocks, resulting in poor quality of covariance estimation (Section \ref{sec:neuro}). At the other end, \cite{creal2024bayesian} choose a prior specification which conditionally shrinks cross-block covariance values to zero, biasing the method toward detecting a small number of large blocks and again resulting in suboptimal performance when the true covariance matrix is better expressed as having many blocks with some negative cross-block covariances. Having adaptive shrinkage requires a prior specification which chooses both the shrinkage target and shrinkage intensity based on the data in a way that the resulting covariance estimate retains accuracy across a variety of true scenarios. 

We introduce such an adaptive method for stochastic block covariance estimation within an empirical Bayesian framework by utilizing a hierarchical prior which promotes an additional layer of shrinkage: conditionally on the partition, the block covariance matrix estimate is biased toward the subspace of homogeneous block matrices. A formal definition is given in Section \ref{sec:shrink}, with the implication that a homogeneous block covariance matrix has the same within and across covariance values. We perform a detailed numerical experiment to show that the ability to shrink toward a sparse subspace -- rather than a single element with block diagonal structure -- and the ability to determine how much to shrink are crucial to offering excellent adaptive performance in estimating block covariance matrices with many blocks and possibly negative cross-block correlations (Section \ref{sec:numerical_prior}). More interestingly, even when the true covariance matrix is not of block structure, the estimation accuracy of our method is comparable to or better than most existing methods, except in a few cases where strong prior information is available on a specific structure of sparsity (Section \ref{sec:numerical_est}). In Section \ref{sec:neuro} we apply the new method to the motivating neuroscience study and identify neuron subpopulations with both positive and negative cross-correlations, some of which are missed by the method of \citep{creal2024bayesian}. In Section \ref{sec:enron}, we revisit the industry classification example studied by \citet{liechty2004bayesian}, but in a more flexible way that does not prefix the number of clusters. In Section \ref{sec:plant}, we employ the proposed method to explore differences in phenotypic integration in plants under contrasting environmental conditions, providing an alternative perspective to the analysis carried out by \cite{matesanz2021phenotypic}. We conclude in Section \ref{sec:discussion} with some remarks on remaining challenges and potential future research directions.

\section{Stochastic Block Covariance Estimation}

We start with the fundamentals of stochastic block covariance model and Bayesian estimation strategies. Much of the discussion in this section is overlapping with \citet{creal2024bayesian} but we add some novel statistical insights such as Theorem \ref{theorem1}. The concepts and notations from this section are going to be useful for our original research on shrinkage presented in the next few sections.

\subsection{Preliminaries of block covariance structure}

We call a $p\times p$ positive semidefinite matrix $\Gamma$ a block covariance matrix if there exist positive integers $k$ and $p_1,\ldots,p_k$ with $p_1 + \cdots + p_k = p$, such that $\Gamma$ can be partitioned as
\begin{equation}
\Gamma=\begin{pmatrix}
\Gamma_{[1,1]} & \Gamma_{[1,2]} & \dots & \Gamma_{[1,k]}\\
\Gamma_{[2,1]} & \Gamma_{[2,2]} & \dots & \Gamma_{[2,k]}\\
\vdots & \vdots &\ddots& \vdots\\
\Gamma_{[k,1]} & \Gamma_{[k,2]} & \dots & \Gamma_{[k,k]}\\
\end{pmatrix},~
\begin{array}{l}
\Gamma_{[u,u]} = (\gamma_u^2 - \gamma_{uu}) I_{p_u} + \gamma_{uu} J_{p_u \times p_u},\\[5pt]
\Gamma_{[u,v]} = \gamma_{uv} J_{p_u\times p_v},~~u\ne v,
\end{array}
\label{equ:bcov}
\end{equation} 
where $I_m$ denotes the $m\times m$ identity matrix, $J_{r\times s}$ denote the $r\times s$ matrix of ones, $\gamma_u^2$ is the common variance of variables in block $u$, $\gamma_{uu}$ is the common within block covariance in block $u$, and $\gamma_{uv}$ is the shared pairwise cross-covariance between blocks $u$ and $v$. Here, $p_u$ denotes the size of group $u$ and $k$ represents the number of blocks. Let $\bb B$ denote the space of all $p\times p$ bloack covariance matrices.

Throughout we assume that data consists of $n$ independent copies $\bm y_1,\ldots, \bm y_n$ of a random vector $\bm y \in \bb R^p$ with $E(\bm y) = 0$ and $\var(\bm y) = \Sigma$ for some positive definite matrix $\Sigma$. We will estimate $\Sigma$ under the Gaussian sampling model
\begin{equation}
    \label{equ:model}
\bm y \sim N(0,\Sigma), ~P\Sigma \in \bb B,~\mbox{for some rotation matrix}~P.
\end{equation}
Any matrix $\Sigma$ satisfying \eqref{equ:model} will be called a {\it group covariance matrix}, i.e., a covariance matrix which could be rotated into a block covariance matrix. Notice that the rotation matrix is not uniquely determined. However, if $\Sigma$ is a group covariance matrix, then there exists a unique and irreducible partition $\sc B = \{B_1, \ldots, B_k\}$ of the indices $\{1,\ldots,p\}$ for some positive integer $k$ such that $\Gamma = \var(P_{\sc B} \bm y)$ is a block covariance matrix with $k$ blocks of sizes $p_u = |B_u|$, where $P_{\sc B}$ is the rotation matrix determined by any permutation $\sigma_{\sc B}$ of $\{1,\ldots,p\}$ such that $\sigma_{\sc B}(B_u) = \{\sum_{v < u} p_v + j: 1 \le j \le p_u\}$, $1 \le u \le k$. Neither $P_{\sc B}$ nor $\Gamma$ is uniquely determined by a group covariance matrix $\Sigma$ but the partition $\sc B$ is. Indeed, $\sc B$ can be recovered from $\Sigma = ((\sigma_{ij}))$ by partitioning $\{1,\ldots,p\}$ according to the equivalence relation: $i \sim j$ if and only if $\sigma_{ii} = \sigma_{jj}$, $\sigma_{il} = \sigma_{jl}$ for every $l \not\in \{i,j\}$. This issue of identifiability will be important in our subsequent treatment.

\subsection{Semi-spectral decomposition of a block covariance matrix}
\label{sec:cr}

For a $\Gamma \in \bb B$ to be positive semidefinite there have to be restrictions on what values $\gamma^2_u, \gamma_{uu}, \gamma_{uv}$ can take. These restrictions can be considerably simplified by using the following semi-spectral factorization \citep{archakov2024canonical} 
\begin{equation}
    \Gamma=QDQ',
    \label{equ:cr}
\end{equation}
where $D$ is a pseudo-diagonal matrix and $Q$ is a $p \times p$ block-sparse orthonormal matrix ($Q'Q=QQ'=I_p$) given as follows:
\[
D\!=\!\!\begin{pmatrix}
A &  &  & \\
 & \!\lambda_1 I_{p_1-1}\! & &\\
& &  \ddots &  \\
 & &  & \!\lambda_k I_{p_k-1}\!
\end{pmatrix}
Q\!=\!\!\begin{pmatrix}
\!Q^{(p_1)}_1\! &  &  & & \!Q^{(p_1)}_{-1}\!& & & \\
& \!Q_1^{(p_2)}\! & & & &\!Q_{-1}^{(p_2)}\!& &\\
&  & \ddots & & & &\ddots&\\
& & &\!Q_1^{(p_k)}\!& & & &\!Q_{-1}^{(p_k)}\!
\end{pmatrix}\!,
\]
with $A = ((a_{uv}))_{k\times k}$ and $\lambda_1, \ldots,\lambda_k$ defined by
\begin{equation}
\label{equ:al}
    \lambda_u=\gamma_u^2-\gamma_{uu},\quad a_{uu}=\gamma_u^2+(p_u-1)\gamma_{uu}, \quad a_{uv}=\gamma_{uv}\sqrt{p_u p_v},~~u\ne v,
\end{equation}
and $Q_1^{(p_u)} = J_{p_u\times 1}/\sqrt{p_u}$ and $Q_{-1}^{(p_u)}$ chosen such that the matrix $Q^{(p_u)}=[Q_1^{(p_u)}~:~Q_{-1}^{(p_u)}]$ are orthonormal matrices.

It is easy to see that $\Gamma$ is positive semidefinite if and only if $A$ is positive semidefinite and $\lambda_1,\ldots,\lambda_k$ are nonnegative. Indeed, if $\bm{z} = (z_1,\ldots,z_p)'$ is a random vector with $\var(\bm{z}) = \Gamma$ then $\var(Q'\bm{z}) = D$. While such a result is reminiscent of the standard spectral decomposition of an arbitrary covariance matrix, a clear advantage of the block assumption is that the orthonormal matrix $Q$ is fully determined by the block sizes alone. Given the blocks, estimation can focus on the $k(k + 3)/2$ many non-zero elements of the matrix $D$ which also has a simple interpretation as the variance matrix of the rotated vector $Q'\bm z$. For example, the non-zero elements of $D$ could be estimated by the corresponding sample quantities of $Q'\bm z$. The next Subsection explores this in more detail. It is worth noting here that $Q$ is not unique except the first $k$ columns which are unique up to sign change. Given block sizes $p_1,\dots,p_k$, one possible version of $Q$ could be constructed using the Gram-Schmidt method; see Examples provided in equations \eqref{equ:Q1} and \eqref{equ:Q2} in the appendix.

\subsection{Estimation of group covariance matrix with known groups}
When the partition $\sc B = \{B_1,\ldots, B_k\}$ is known, the observation vector $\bm y$ can be rotated to $\bm z = P_{\sc B} \bm y$ and the observation model could be rewritten as $\bm z \sim N(0, \Gamma)$, $\Gamma \in \bb B$, with $p_u = |B_u|$, $1 \le u \le k$. The semi-spectral decomposition result then gives $\Gamma = QDQ'$ with $Q$ being a known orthogonal matrix and $D$ an unknown block diagonal matrix determined by a $k \times k$ positive semidefinite matrix $A$ and $k$ nonnegative numbers $\lambda_1,\ldots,\lambda_k$. The laws of large numbers imply that $\hat D$, which estimates the non-zero elements of $D$ by the corresponding sample quantities of $Q'\bm z$, gives a consistent estimate of $D$ as sample size $n \to \infty$ for a fixed dimension $p$. The estimate $\hat D$ is also the maximum likelihood estimate of $D$ under the Gaussian model. 

Given the known partition $\sc B = (B_1,\ldots,B_k)$ take $\bm z_i = P_{\sc B} \bm y_i$, $i = 1,\ldots,n$. Find $Q$ in the semi-spectral decomposition, which only depends on the total number of blocks $k$ and the block sizes $p_1,\ldots,p_k$. Define $\bm\eta_i = Q'\bm z_i$. Partition each rotated vector $\bm\eta_i$ as $\bm{\eta}_i'=(\bm{\eta}_{i(0)}',\bm{\eta}_{i(1)}',\ldots,\bm{\eta}_{i(k)}')$ where $\bm\eta_{i(0)}$ is of length $k$ and $\bm\eta_{i(u)}$ is of length $p_u - 1$, $u = 1,\ldots, k$. The log-likelihood function can be written as
\begin{equation}
\label{eq:loglik}
    \text{loglik}~\dot=
    -\tfrac{n}{2}\log |A|-\tfrac{1}{2}\sum_{i=1}^n \bm{\eta}_{i(0)}'A^{-1}\bm{\eta}_{i(0)}-\tfrac{n}{2}\sum_{u=1}^k (p_u-1)\log \lambda_u -\tfrac{1}{2}\sum_{i=1}^n\sum_{u=0}^k \frac{\|\bm{\eta}_{i(u)}\|^2}{\lambda_u},
\end{equation}
    where $\dot=$ denotes equality up to an additive constant. By taking the derivatives of the log-likelihood with respect to $A$ and the $\lambda_k$'s respectively, we can obtain the maximum likelihood estimator 
\begin{equation}
\label{equ:mle}
    \hat{A}=\frac{\sum_{i=1}^n\bm{\eta}_{i(0)}\bm{\eta}_{i(0)}'}{n}, \quad \hat{\lambda}_u=\frac{\sum_{i=1}^n\|\bm{\eta}_{i(u)}\|^2}{n(p_u-1)}
\end{equation}
which are precisely the non-zero elements of $\hat D$. 

The above result about $\hat D$ is presented in \cite{archakov2024canonical} (see Theorem 3), but they do not explore properties of the corresponding estimate $\hat \Gamma = Q'\hat D Q$ of $\Gamma$ or that of $\hat \Sigma = P_{\sc B}'\hat \Gamma P_{\sc B}$. The theorem below, which seems to be new to the literature, shows that $\hat \Sigma$ could be computed simply by replacing the blocks of the sample covariance matrix $S = \sum_i \bm y_i \bm y_i'/n$ with block level averages.

\begin{theorem}
\label{theorem1}
$\hat \Gamma = Q \hat D Q'$ has the same block structure \eqref{equ:bcov} as of $\Gamma$ with $\gamma^2_u$, $\gamma_{uu}$, $\gamma_{uv}$ replaced with the following estimates
\[
\hat \gamma^2_u = \frac{\sum_{j \in B_u} S_{jj}}{p_u}, \quad \hat \gamma_{uu} = \frac{\sum_{j  \ne j' \in B_u} S_{jj'}}{p_u(p_u-1)}
, \quad \hat \gamma_{uv} = \frac{\sum_{j \in B_u, j'\in B_v} S_{jj'}}{p_up_v}~\mbox{if}~u\ne v.
\]
Accordingly, the elements of $\hat \Sigma = ((\hat \sigma_{ij}))$ are
\[
\hat \sigma_{ij} =\left\{\begin{array}{ll} 
\frac{\sum_{l \in B_u} S_{ll}}{p_u}& \mbox{if}~i = j \in B_u\\
\frac{\sum_{l  \ne l' \in B_u} S_{ll'}}{p_u(p_u-1)} & \mbox{if}~i \neq j\in B_u,\\\frac{\sum_{l \in B_u, l'\in B_v} S_{ll'}}{p_up_v} & \mbox{if}~i\in B_u,j\in B_v, u\ne v.
\end{array}\right.
\]
\end{theorem}
A proof is presented in Appendix \ref{append:b}. Notice again that $\hat \Sigma$ is free of ambiguity regarding $P_{\sc B}$ or $\Gamma$. At this point it might be tempting to think that the maximum likelihood framework could be extended also learn about the blocks $\sc B$ by considering the profile loglikelihood $L^*(\sc B) = \max_{\Sigma} p(Y|\Sigma, \sc B) \propto |\hat A|^{-n/2}\prod_{u = 1}^k \hat \lambda_u^{-n(p_u - 1)2}$. However, estimation of $\sc B$ based on $L^*(\sc B)$ suffers massively from the curse of dimensionality -- the profile likelihood favors putting each variable in its own cluster with $k \approx p$ and $\hat \Sigma \approx S$, and fails to offer any useful regularization (not reported here). A much more useful framework could be constructed by turning to a Bayesian formulation which offers natural protections against the curse of dimensionality by replacing the profile likelihood with a marginal likelihood obtained by integrating over $p(Y|\Sigma, \sc B)$ with respect to a prior $p(\Sigma | \sc B)$  \citep{jefferys1992ockham}. It turns out that for the current problem, a rich class of conjugate prior distributions are available for $p(\Sigma | \sc B)$, which not only leads to an easy estimation of $\Sigma$ given $\sc B$, but also an easy calculation of the marginal likelihood $p(Y|\sc B) = \int p(Y|\Sigma, \sc B)p(\Sigma | \sc B)d\Sigma$. This is what we discuss next.

\subsection{Conjugate prior under known groups} 
A quick inspection of the log-likelihood formula \eqref{eq:loglik} reveals that when the grouping structure $\sc B = \{B_1,\ldots, B_k\}$ is known with $p_u = |B_u|$, $1 \le u \le k$, the specification
\begin{equation}
\label{equ:prior_plain}
\textstyle  (A,\lambda_1,\ldots, \lambda_k) \mid \sc B \sim IW(\nu_0+k+1,\nu_0 A_0) \times \otimes_{u=1}^k IG(\frac{s_{0,u}+2}2, \frac{s_{0,u}\lambda_{0,u}}2),
\end{equation}
gives a conjugate prior distribution for $A$ and $\lambda_1, \ldots,\lambda_u$, with the posterior distribution given $Y = (\bm y_1,\ldots,\bm y_n)$ equaling
\begin{equation}
\label{equ:A}
\textstyle    (A,\lambda_1,\ldots, \lambda_k) \mid (Y, \sc B) \sim IW(\nu_n+k+1,\nu_n A_n) \times \otimes_{u=1}^k IG(\frac{s_{n,u}+2}2, \frac{s_{n,u}\lambda_{n,u}}2),
\end{equation}
with the usual updates
\begin{align*}
&\nu_n = \nu_0 + n,~~ \nu_nA_n = \nu_0A_0 + n\hat A,\\ 
&s_{n,u} = s_{0,u} + n(p_u-1), ~~ s_{n,u}\lambda_{n,u} = s_{0,u}\lambda_{0,u} + n(p_u-1)\hat \lambda_u
\end{align*}
where $\hat A$ and $\hat \lambda_u$ are as before. 

As an estimate of $\Sigma$ one could take the posterior mean $\Sigma_n = E(\Sigma | Y,\sc B) = P_{\sc B}'\Gamma_n P_{\sc B}$ with $\Gamma_n = QD_n Q'$ where $D_n$ is block diagonal with blocks $(A_n, \lambda_{n,1} I_{p_1-1}, \ldots, \lambda_{n,K} I_{p_K-1})$. Clearly, 
\[
\Sigma_n = \alpha_n \Sigma_0 + (1 - \alpha_n) \hat \Sigma
\]
where $\alpha_n = \nu_0/(\nu_0 + n) \in (0,1)$, $\hat \Sigma$ is as before, and $\Sigma_0 = P_{\sc B}' \Gamma_0 P_{\sc B}$ with $\Gamma_0 = Q D_0 Q'$ where $D_0$ is the block diagonal matrix with blocks $(A_0,\lambda_{0,1}I_{p_1-1},\ldots,\lambda_{0,k} I_{p_k - 1})$. An interesting choice of the hyperparameters is the data dependent choice: $\nu_0 = g^{-1}n$, $s_{0,u} = g^{-1}n(p_u - 1)$, $A_0 = \hat A$, $\lambda_{0,u} = \hat \lambda_u$, which gives $\Sigma_n =  \Sigma_0 = \hat\Sigma$. We will refer to this as the hyper-inverse Wishart $g$-prior following by \cite{carvalho2009objective} who used similar constructions for graph learning. 

Conjugacy is salient in the present context on two accounts. It shows that a prior distribution can be constructed easily and directly on the space of a group covariance matrices with known grouping structure, and thus, offering significant improvement over the ad hoc and truncation-based construction of \cite{liechty2004bayesian}. Second, and this point is crucial, a marginal likelihood for the block partition $\sc B = \{B_1,\ldots,B_k\}$ can be calculated by integrating out $A$ and $\lambda_1,\ldots,\lambda_k$:
\begin{equation}
\label{equ:mlik}
p(Y|\sc B) = (2\pi)^{-\frac{np}2}\times 2^{\frac{nk}{2}}\times\frac{|\nu_0A_0|^{\frac{\nu_0}2}{\Gamma_k (\frac{\nu_n}{2})}}{|\nu_nA_n|^{\frac{\nu_n}2}{\Gamma_k (\frac{\nu_0}2)}}   \prod_{u=1}^k\frac{\Gamma(\frac{s_{n,u}}{2})(\frac{s_0\lambda_0}{2})^{\frac{s_0}{2}}}{\Gamma(\frac{s_0}{2})(\frac{s_{n,u}\lambda_{n,u}}{2})^{\frac{s_{n,u}}{2}}}
\end{equation}
Note that the marginal likelihood $p(Y|\sc B)$ is free of any issues of identifiability, because $\sc B$ is uniquely determined by a group covariance matrix $\Sigma$. We discuss next how this likelihood function could be combined with an appropriate prior on $\sc B$ to carry out posterior inference on the blocks. 


\subsection{Estimation of blocks under partition priors}
\label{sec:B-prior}
For our purposes it would be desirable for a prior distribution on $\sc B$ to be invariant under permutation of the variable indices $1,\ldots,p$. Probability distributions satisfying this permutation invariance property are referred to as exchangeable partition probability functions (EPPF) as introduced by \cite{kingman1978representation} and refined thereafter by many other authors; see \cite{pitman2006combinatorial} for a review. EPPFs have a close connection to complete random measures which are routinely used in Bayesian mixture models, with the Dirichlet process mixtures \citep{ferguson1973bayesian, antoniak1974mixtures} and its generalizations \citep{pitman1997two} being the most widely known examples. In this article we work with the {\it mixture of finite mixtures} (MFM) partition distribution \citep{miller2018mixture} which, compared to Dirichlet process mixtures, offers a greater control on the number of blocks and favors partitions with blocks of similar sizes. Its EPPF is given as
\begin{equation}
    \label{equ:eppf}
p(\sc B) = p(B_1,\ldots,B_k) = V_p(k)\prod_{u = 1}^k \rho^{(|B_u|)}
\end{equation}
with $V_p(k) = \sum_{u = 1}^\infty f(u)u_{(k)}/(\rho u)^{(p)}$ where $x^{(k)}:=x(x+1)\cdots(x + k-1)$ and $x_{(k)} = x(x-1)\cdots(x-k+1)$, and $f(k)$ is any strictly positive probability mass function on $\{1,2,\dots\}$ such that $\sum_{k=1}^{\infty} f(k)$ converge to 1 reasonably quickly. The parameter $\rho$ is a type of concentration parameter. Larger values of $\rho$ favor blocks of more equal size. 

By combining such a prior on $\sc B$ with the conjugate prior in \eqref{equ:prior_plain} and the marginal likelihood $p(Y|\sc B)$ from earlier, one obtains the posterior distribution of $\sc B$ as $p(\sc B|Y) \propto p(\sc B)p(Y | \sc B)$, which could be used to estimate $\Sigma$ as $\bar \Sigma := E(\Sigma | Y) = \sum_{\sc B} E(\Sigma | Y, \sc B)p(\sc B | Y)$, with $E(\Sigma | Y,\sc B)$ as described before. As is typical in Bayesian estimation, $\bar\Sigma$ could be approximated by Markov chain Monte Carlo (MCMC) as $\bar \Sigma \approx(1/ T) \sum_{t = 1}^T E(\Sigma | \sc B = \sc B^{(t)})$, where $(\sc B^{(t)}, t = 1,2,\ldots)$ is an ergodic Markov chain with $p(\sc B |Y)$ as its stationary distribution. 

For the MFM prior, such a Markov chain can be easily constructed and sampled from by using Gibbs sampling. Indeed, if we choose $f(k)$ to be the unit mean Poisson distribution shifted to the right by 1, then 
%
%
%
a random element $\sc B$ drawn from the MFM EPPF can be described as follows
\begin{equation}
    \begin{split}
            k^*\sim f \\
(\pi_1,\dots,\pi_{k^*})\sim \text{Dir}_{k^*}(\rho,\dots,\rho)\\
c_1,\dots,c_p \sim \text{Mult}(\pi)\\
\sc B = \text{partition}(i \sim j~\text{iff}~c_i = c_j)
    \end{split}
    \label{equ:mfm_eppf}
\end{equation}
where the partition operation in the last step simply divides the indices $1,\ldots,p$ into disjoint non-empty subsets according to ties in the corresponding labels $c_1,\ldots,c_p$ generated in the previous step. Note that $\sc B = \{B_1, \ldots, B_k\}$ for some $1 \le k \le k^*$. 
A Gibbs chain with $p(\sc B|Y)$ as its stationary distribution can be constructed as follows; see \citet{miller2018mixture} for technical details.

\begin{enumerate}
    \item Initialize $\mathcal{B}^{(0)} = \{\sc B^{(0)}_1,\ldots,\sc B^{(0)}_k\}$ as an arbitrary partition of $\{1,\ldots,p\}$ and create labels $c_1,\ldots,c_p$ so that $c_i = c_j$ if and only if $i$ and $j$ belong to the same partition element $\sc B_u$ for some $u \in\{1,\ldots,k\}$. For example, one could take $\sc B^{(0)} = \{\{1,\ldots,p\}\}$, $k = 1$, $c_1 = \cdots = c_p = 1$.
    \item For $t = 2,\ldots,T$, repeat:
    \begin{enumerate}
        \item 
        For $i=1,\dots,n$ repeat:
        \begin{enumerate}
            \item Let $\sc B^{-} = (B^-_1,\ldots,B^-_{k^-}\}$  denote the partition of $\{1,\ldots,p\} \setminus \{i\}$ determined by the ties in the labels $\{c_1,\ldots,c_p\}\setminus \{c_i\}$ and let $c^*_1,\ldots,c^*_{k^-}$ be the unique labels among $\{c_1,\ldots,c_p\}\setminus \{c_i\}$. For each $u \in \{1,\ldots,k^-\}$ let $\sc B^{*u}$ denote the extension of $\sc B^-$ to a size $k^-$ partition of $\{1,\ldots,p\}$ obtained by adding $i$ to the subset $B^-_u$. Also let $\sc B^{**} = \sc B^- \cup \{i\}$ denote a size $(k^- + 1)$ partition of $\{1,\ldots,p\}$ obtained by extending $\sc B^-$ where $i$ is placed in a singleton subset of its own. 
            \item Draw a new label $c_i$ for index $i$ either from $\{c^*_1,\ldots,c^*_{k^-}\}\setminus \{c_i\}$ or as a completely new and arbitrary value $c^*$ as follows:
            \[
            \Pr(c_i = c) \propto \left\{\begin{array}{ll} (|B^-_u| + \rho)\times p(Y|\sc B^{*u}) &~\mbox{if}~c = c^*_u,~u \in \{1,\ldots,k^-\}\\[5pt] \frac{V_n(k^-+1)}{V_n(k^-)} \times p(Y|\sc B^{**})& ~\mbox{if}~c=c^*\not\in\{c^*_1,\ldots,c^*_{k^-}\}\end{array}\right.
            \]
        \end{enumerate}
        
        \item Set $\sc B^{(t)} = \text{partition}(i \sim j~\text{iff}~c_i = c_j)$.
        
    \end{enumerate}

\end{enumerate}
These Gibbs updates are similar to those seen for typical nonparametric Bayesian mixture models, with the important difference that calculation of the label reassignment probabilities involves evaluation of the entire marginal likelihood $p(Y| \sc B)$ for $k^- + 1$ possible assignments of $c_i$. These evaluations can be time consuming unless some care is taken to streamline calculations. For example, it would be helpful to parallelize the $k^-+1$ calculations to multiple processors. Given the expectation that $k^-$ would typically be quite modest, such parallelization can be accomplished on modern multicore personal computers.

The Gibbs sampler described above incrementally updates one label $c_i$ keeping all others fixed. While such incremental updating is common to the nonparametric mixture literature, it has known limitations such as difficulty with breaking apart a prematurely formed large block. The stickiness of large blocks leads to slow mixing as one needs a large number of iterations to chip away misfit members of the large block one by one and reassemble them into a separate block. This problem is more pronounced for the MFM prior which typically discourages formation of small blocks and thus becomes more sticky when a large block forms prematurely. Merge-split samplers can overcome this problem by allowing randomly splitting one existing block into two or merging a pair \citep{dahl2003improved, dahl2005sequentially, jain2004split, jain2007split}. 

Traditional merge-split samplers assign items independently and with equal probability during the split process. \citet{dahl2022sequentially} introduced an adaptive procedure, known as the Sequentially Allocated Merge-Split Sampler (SAMS), which splits a large block into two by starting with a random pair as seeds for the new sub-blocks and then allocating remaining items sequentially conditional on previously allocated items. We adopt the SAMS method in our posterior sampling, where a full Gibbs scan as described earlier is interpresed with five repeats of SAMS.

\section{Hyperparameter Learning and Shrinkage}
\label{sec:shrink}
So far we have not discussed the choice of the hyperparameters in the conjugate prior specifications $A \sim IW(\nu_0 + k + 1,A_0)$ and $\lambda_u \sim IG((s_{0,u}+2)/2,s_{0,u}\lambda_{0,u}/2)$. In absence of strong prior beliefs about model parameters, Bayesians often adopt a weakly informative prior which reduces estimation variance by inducing mild estimation bias toward a reasonable prior mean. For the inverse-Wishart part of the prior, it is quite commonplace in the literature to adopt a modest degree of freedom and a scaled identity matrix as its mean. In our context, a specific choice along this line would be
\begin{equation}
\label{prior:wi}
        \nu_0=2,\quad A_0=a_{00} I_k, \quad s_{0,u}=2, \quad \lambda_{0,u}=\tau,
\end{equation}
where $a_{00} > 0$ is chosen in a data dependent way so that the prior mean matches the scale of the data. 

Is the weakly-informative choice \eqref{prior:wi} any good? We show below and also in Section \ref{sec:experiments} that the posterior $p(\sc B | Y)$ is quite sensitive to the choice of the hyperparameters which, as can be deduced from \eqref{eq:loglik}, exert a direct and great influence on the marginal likelihood $p(Y | \sc B)$. Recall that $p(Y|\sc B)$ equals an integral of the likelihood function $p(Y|\Gamma, \sc B)$ over a $k(k+3)/2$ dimensional space with respect to the conditional prior $p(\Gamma | \sc B)$. Consequently, its evaluation critically depends on how the centering and the concentration of this prior distribution changes with $k$. We contend that for good estimation of $\sc B$ one needs a flexible hierarhical specification so that the hyperparameters in \eqref{equ:prior_plain} which determine the centering and the concentration of $p(\Gamma | \sc B)$ can be judiciously adjusted across various choices of $\sc B$ in a data-dependent manner. A thorough examination of this issue is what makes our work fundamentally different from the treatment presented in \cite{creal2024bayesian}.

To gain a deeper understanding of the role of the prior means $A_0$ and $\lambda_{0,1},\ldots,\lambda_{0,k}$, it helps to look at the corresponding prior mean $\Gamma_0$ of $\Gamma$ (given a blocking structure $\sc B$). From \eqref{equ:al}, $\Gamma_0$ is a block matrix with $k$ blocks of sizes $p_1,\ldots,p_k$, determined by the unique elements 
\[
\gamma^2_{0,v}=  \frac{a_{0,uu} + (p_u - 1) \lambda_{0,u}}{p_u}, \quad \gamma_{0,uu} =\frac{a_{0,uu} - \lambda_{0,u}}{p_u}, \quad \gamma_{0,uv} =  \frac{a_{0,uv}}{\sqrt{p_up_v}},~u\ne v.
\]
These relations make it clear that picking a prior mean $\Gamma_0$ may not be easy or even meaningful if one only paid attention to $A_0$ and $\lambda_{0,u}$, without taking into account the block structure $\sc B$, especially when the blocks are of different size from one another. Seen through this lens, the weakly-informative prior in \eqref{prior:wi} does not appear a compelling choice in producing a reasonable target value $\Gamma_0$ for $\Gamma$. Indeed, it gives $\Sigma_0 = \Gamma_0 = a_{00} I_p$, which completely ignores the block structure of the associated $\sc B$.

To incorporate the information in $\sc B$ into the prior mean $\Gamma_0$, \citet{creal2024bayesian} recommend setting reasonable values directly for $\Gamma_0$. Their exact recommendation is slightly complicated, but a nearly equivalent but simplified form would be to set $\gamma^2_{0,u} \equiv \tau_0$, $\gamma_{0,uu} = \tau_0 r_0$, and $\gamma_{0,uv} = 0$, $u\ne v$, where $\tau_0 = \text{median}(S_{11},\ldots,S_{pp})$ and $r_0 \ge 0$ is a target value for within-block correlations $r_{uu} := \gamma_{uu}/\gamma^2_u$. These choices translate to the following choices of the hyperparameters:
\begin{equation}
\label{prior:ck}
A_0 = \tau_0 \times \text{diag}(1+r_0(p_1-1),\ldots,1+r_0(p_k-1)),~\lambda_{0,u} = (1-r_0)\tau_0.    
\end{equation}
\citet{creal2024bayesian} recommend setting the value of $r_0$ based on the context of the study and adopt $r_0 = 0.35$ for their experiments and applications. Clearly, any such choice of $r_0$ is rather ad hoc. In any given application it would be difficult to determine if $r_0 = 0.2$ or $r_0 = 0.5$ would have made a better choice than $r_0 = 0.35$. Notice that the choices described in \eqref{prior:wi} also match \eqref{prior:ck} with $r_0 = 0$, and hence \eqref{prior:wi} could be taken as a variant of the \citet{creal2024bayesian} prior specification. \citet{creal2024bayesian} acknowledge the difficulty in specifying a well reasoned value for $r_0$, and hence recommend taking the precision parameters $\nu_0$ and $s_{0,u}$ to be modest, both restricted to the integers between 1 and 10, so that the ad hoc choice of $r_0$ could be easily washed out by the information in the data.

\begin{figure}[!t]
\centering
\includegraphics[scale=1]{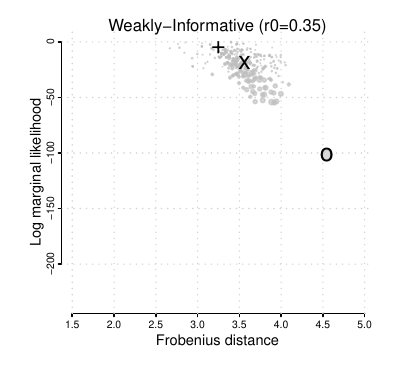}
\includegraphics[scale=1]{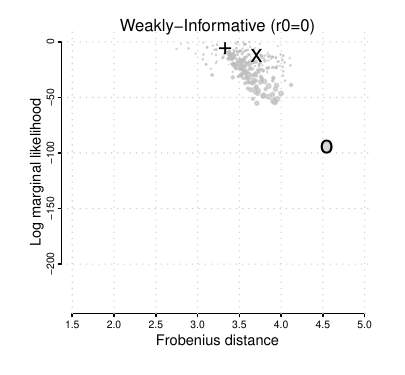}
\includegraphics[scale=1]{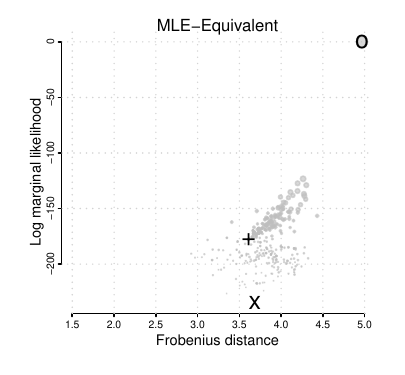}
\includegraphics[scale=1]{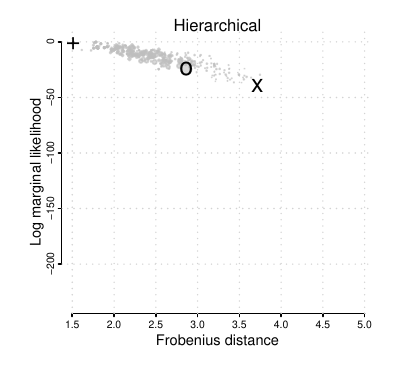}
\caption{The effect of hyperparameter choice on the inference on $\sc B$ and $\Sigma$. One synthetic dataset of size $n = p = 20$ was generated under the group covariance matrix model assumption with $k = 7$ and block sizes $p_1 = 9, p_2 = 6$, and $p_3 = \cdots = p_7 = 1$. The marginal likelihood $p(Y|\sc B)$ (vertical axis, log scale) and the Frobenius distance (horizontal axis) between the true $\Sigma$ and the posterior mean $\Sigma_n = E(\Sigma | Y, \sc B)$ were calculated for 200 small variations of the true partition $\sc B$, each variant is represented by one gray dot with the size of the dot  indicating the size of the associated $\sc B$. The point marked "+" shows the true data generating partition $\sc B$, the point marked "o" shows the cluster of size 20 where each block is of size 1, and the point marked "x" shows the cluster with one block of size 20. The weakly-informative priors can be seen to favor small number of blocks even when estimation accuracy of $\Sigma$ is relatively the same, whereas the MLE-equivalent prior clearly favors as many blocks as the number of variables. The hierarchical prior not only improves the estimation quality of $\Sigma$ for most $\sc B$, it also offers a more direct relationship between the quality of $\hat \Sigma$ and the correspoinding $p(Y|\sc B)$.}
\label{fig:prior}
\end{figure}

We demonstrate here and also in Section \ref{sec:experiments} that such weakly informative priors may lead to underestimation of the block count $k$, thus leading to a large bias in the estimation of $\Sigma$ in spite of the appearance of being weakly-informative for the conditional prior specifications \eqref{equ:prior_plain} given $\sc B$. This bias is potentially due to the Occam's razor property of Bayes factors \citep{jefferys1992ockham}. Recall again that the marginal likelihood $p(Y | \sc B)$ equals the integral of $p(Y|\Gamma,\sc B)$ over the $k(k+3)/2$ free elements of $\Gamma$ with respect to the prior on $A$ and $\lambda_1,\ldots,\lambda_k$ where $k = |\sc B|$. With modest values for $\nu_0$ and $s_{0,u}$, the larger the $k$, the smaller this integral would tend to be because the prior distribution will be more diffuse in higher dimensional spaces, even if the estimation accuracy of the corresponding posterior mean $\Sigma_n$ did not get any worse. This phenomenon is illustrated in Figure \ref{fig:prior} which shows the relationship between $p(Y|\sc B)$, size of $\sc B$, and the estimation accuracy of $\Sigma_n = E(\Sigma |Y, \sc B)$ for one synthetic data set. It is also evident from the same figure that the performance of the weakly-informative prior remains about the same between the choices of $r_0 = 0.35$ as in \cite{creal2024bayesian} and $r_0 = 0$ as in \eqref{prior:wi}. In other words, when the prior is weakly informative about $A$ and $\lambda_{u}$, the exact information does not have a substantive impact on estimation.

An alternative to specifying weakly-informative priors is to extract some information from the data and use it to construct a data-dependent prior. An attractive choice of this type is the empirical $g$-prior $p(A,\lambda_1,\ldots,\lambda_k \mid \sc B) \propto p(Y \mid A, \lambda_1,\ldots,\lambda_k,\sc B)^g$ for a small value of $g$, e.g., $g = 1/n$. \cite{carvalho2009objective} use a similar $g$-prior in a graph selection problem where it appears to offer useful regularization in sparse graph learning from limited samples. For our model, the resulting $g$-prior distribution is of the conjugate form as in \eqref{equ:prior_plain} with
\begin{equation}
\label{prior:mle}
    \nu_0=1, \quad A_0=\frac{\sum_{i=1}^n\bm{\eta}_{i(0)}\bm{\eta}'_{i(0)}}{n}, \quad s_{0,u} = p_u - 1, \quad \lambda_{0,u} = \frac{\sum_{i = 1}^n \|\bm{\eta_{i(u)}}\|^2}{n(p_u-1)}.
\end{equation} 
Instead of fixing arbitrary prior means for $A$ and $\lambda_u$, the $g$-prior sets the prior means in a data dependent manner. We call it an MLE-equivalent prior because the corresponding posterior mean $E(\Sigma | Y, \sc B)$ is precisely the MLE $\hat \Sigma$ described in Theorem \ref{theorem1}. While the MLE-equivalent prior avoids introducing bias in the estimation of $\Sigma$ given $\sc B$, Figure \ref{fig:prior} suggests it favors complex partitions, leading to severe overestimation of the number of blocks. This behavior is unsurprising because the MLE-equivalent prior clearly fails to offer any regularization against overfitting.

A more fruitful way to construct a data dependent prior is to consider an empirical or hierarchical Bayesian formulation where, instead of a single prior choice, one considers an entire family of prior distributions indexed by a small number of parameters to be estimated from the data. The advantage of such a construction is that, once these higher level parameters are estimated, calculations proceed with the corresponding element of the prior family to be used as the prior for the original analysis. The selected prior could help reduce bias as well as provide regularization much more effectively than any single prefixed prior distribution. See \citet{efron2019bayes} for a thorough discussion. 

For our analysis, a hierarchical framework incorporating a family of prior distributions could be constructed by considering all $\Gamma_0$ of the form
\[
\gamma^2_{0,u} \equiv \gamma^2_{0,\text{variance}}, \quad \gamma_{0,uu} \equiv \gamma_{0,\text{within}}, \quad \gamma_{0,uv} \equiv \gamma_{0,\text{between}},~u\ne v,
\]
determined by three unknown scalar quantities $\gamma^2_{0,\text{variance}} \ge \gamma_{0,\text{within}} \ge \gamma_{0,\text{between}}$. By \eqref{equ:al} this immediately fixes $A_0$ and $\lambda_{0,u}$ to be
\[
    \lambda_{0,u}=\gamma_{0,\text{variance}}^2-\gamma_{0,\text{within}},~~ a_{0,uu}=\gamma_{0,\text{variance}}^2+(p_u-1)\gamma_{0,\text{within}}, ~~ a_{0,uv}=\gamma_{0,\text{between}}\sqrt{p_u p_v},~u\ne v.
\]
In other words, $(A_0, \lambda_{0,1},\ldots,\lambda_{0,k})$ reside within a 3-dimensional subspace such that the prior mean of $\Gamma$ is a homogeneous block covariance matrix where all block-specific variances are the same, all within-block covariances are the same, and all cross-block covariances are the same. Note that the weakly-informative choice \eqref{prior:ck} of \cite{creal2024bayesian} is one point within this homogeneous subspace with $\delta_1 = \tau_0 (1-r_0)$, $\delta_2 = 0$, $\delta_3 = \tau_0r_0$. Having a homogeneous block matrix as the prior mean allows information sharing across blocks, where block-level variances and covariances for smaller blocks are shrunk more toward central values which are estimated from the data and are likely influenced more heavily by information from larger blocks. Seen through this lens, the hierarchical prior could be seen as a facilitator of estimation shrinkage -- with the shrinkage target and the shrinkage intensity both learned from the data. 

Toward estimating the shrinkage target and intensity, we take $s_{0,u} \equiv s_0$, and reparametrize $\delta_1 = \gamma^2_{0,\text{variance}} - \gamma_{0,\text{within}}$, $\delta_2 = \gamma_{0,\text{between}}$, $\delta_3 = \gamma_{0,\text{within}} - \gamma_{0,\text{between}}$, so that
\begin{equation}
    \lambda_{0,u}=\delta_1,~~a_{0,uv}=\sqrt{p_up_v}\delta_2 + 1_{u = v} (\delta_1 + p_u\delta_3),
    \label{prior:hi3}
\end{equation}
and specify the hyper-priors $\delta_1 \sim Ga(2,4)$, $\delta_2 \sim Ga(10,1)$, $\delta_3 \sim Ga(10,1)$. We also assign priors to control shrinkage intensity as $\log(\nu_0 - 2) \sim T_1(0,1)$, $\log s_0 \sim T_1(0,1)$, allowing a broad range of shrinkage intensities. Posterior estimation of $(\nu_0,s_0, \delta_1, \delta_2, \delta_3)$ can be easily performed by expanding the Markov chain Monte Carlo method described earlier; see Appendix \ref{sampler:expand-mcmc}. 

Figure \ref{fig:prior} suggests that the hierarchical specification could be far superior to the weakly informative priors or the MLE-equivalent prior for estimating both $\sc B$ and $\Sigma$. The hierercahical prior learns both the prior mean and the prior concentration for $A$ and $\lambda_{u}$ (given $\sc B$) from the data and leads to much better estimates of $\Sigma$ than the other choices. At the same time, it assigns lower marginal likelihood score $p(Y|\sc B)$ to partitions $\sc B$ for which the associated estimate $\Sigma_n$ is not very good. Indeed, unlike the weakly-informative or MLE-equivalent specifications, under the hierarchical setting, the magnitude of $p(Y|\sc B)$ seems much more directly related to the quality of $\Sigma_n$ rather than simplistic features such as the size of $\sc B$. We show in the next section that the favorable performance of the hierarhical specification is quite typical across various data generating processes and that the resulting estimation method is an excellent candidate for high-dimensional covariance matrix estimation even when the group covariance assumption is somewhat misspecified.

\section{Numerical Experiments}
\label{sec:experiments}
Here we report results from numerical experiments we ran to examine two key questions. When the group covariance assumption is reasonably accurate, what advantages are gained with the hierarchical prior which attempts to adaptively shrink toward a subspace of homogeneous block matrices? How good is the quality of covariance estimation when the group covariance assumption is false? In answering the first question, we compared the hierarchical prior against the weakly-informative prior of \eqref{prior:wi} which is a special case of \eqref{prior:ck} with $r_0 = 0$ and $\nu_0$ and $s_0$ fixed at 2. Performance was assessed according to both statistical accuracy of the covariance estimate and correct recovery of the partition $\sc B$.  For the second question, we examined the covariance estimation accuracy under various alternative sparsity assumptions or other lower-dimensional structures, including block-diagonal structures, order-dependent sparsity, entrywise-independent sparsity, block-sparsity, and factor-based sparsity. For these evaluations, we compared our method against other bespoke methods such as banding, tapering, threshold-based, and Ledoit-Wolf shrinkage estimators.

\subsection{Impact of shrinkage on block covariance estimation}
\label{sec:numerical_prior}
In our first experiment, we examined the relative performance of the hierarchical prior against the weakly-informative prior when the group covariance matrix assumption was correct. 
In generating synthetic data, we fixed the dimension $p = 50$ and considered two possible samples sizes: $n \in \{25, 50\}$. Results from $n = 25$ are presented here. See Figure \ref{fig:well.n50.fn} and Figure \ref{fig:well.n50.ari} in the Appendix for $n=50$. We considered three possible maximum cluster counts $k^* \in \{5,10,20\}$. For each combination of $(p,n,k^*)$, we considered $3\times 3$ scenarios which differed from one another in terms of how the true grouped covariance matrix $\Sigma$ was set. For each scenario, 100 replicates of $(\sc B, \Sigma, Y)$ were generated as follows. We first generated $\sc B$ by randomly assigning each variable $i \in \{1,\ldots,p\}$ a label $c_i \in \{1,\ldots,k^*\}$ with $\Pr(c_i = u) \propto \max(0.1,0.7^u)$. This ensured that the actual cluster count $k \le k^*$ and the clusters were of unequal relative sizes. Next we generated $\Sigma = P_{\sc B}\Gamma$, where $\Gamma$ was a block covariance matrix with $k$ blocks which was generated according to the prior \eqref{equ:prior_plain} with $\nu_0$ and all $s_{0,u}$ assigned a common value $\tau \in \{1,10,100\}$ and $A_0$ and $\lambda_{0,u}$ set according to the hierarchical specification \eqref{prior:hi3} with $(\delta_1,\delta_2,\delta_3)$ taken to be one of the three possible configurations: 
\begin{itemize}
    \item $\delta_1=0.5,\delta_2=0,\delta_3=0$ (diagonal $\Gamma_0$),
    \item $\delta_1=0.5,\delta_2=0,\delta_3=0.5$ (block-diagonal $\Gamma_0$), or 
    \item $\delta_1=0.5,\delta_2=0.2,\delta_3=0.3$ (homogeneous block $\Gamma_0$). 
\end{itemize}
Finally, $Y = (\bm y_1,\ldots,\bm y_n)$ was generated as $\bm y_i \sim N(0,\Sigma)$. Note that the three choices of $\tau$ and three choices of $(\delta_1,\delta_2,\delta_3)$ give rise to the $3\times 3$ sub-design within each $(p,n,k^*)$ configuration of our experimental design. When the precision parameter $\tau$ is large, the true covariance matrix $\Sigma$ is nearly a homogeneous grouped covariance matrix. But when $\tau = 1$ or 10, $\Sigma$ could have fairly inhomogeneous within- and cross-block covariances as well as varying block variances.

\begin{figure}[!t]
\begin{center}
\includegraphics[scale=0.43,trim=0 0 3.5cm 0,clip]{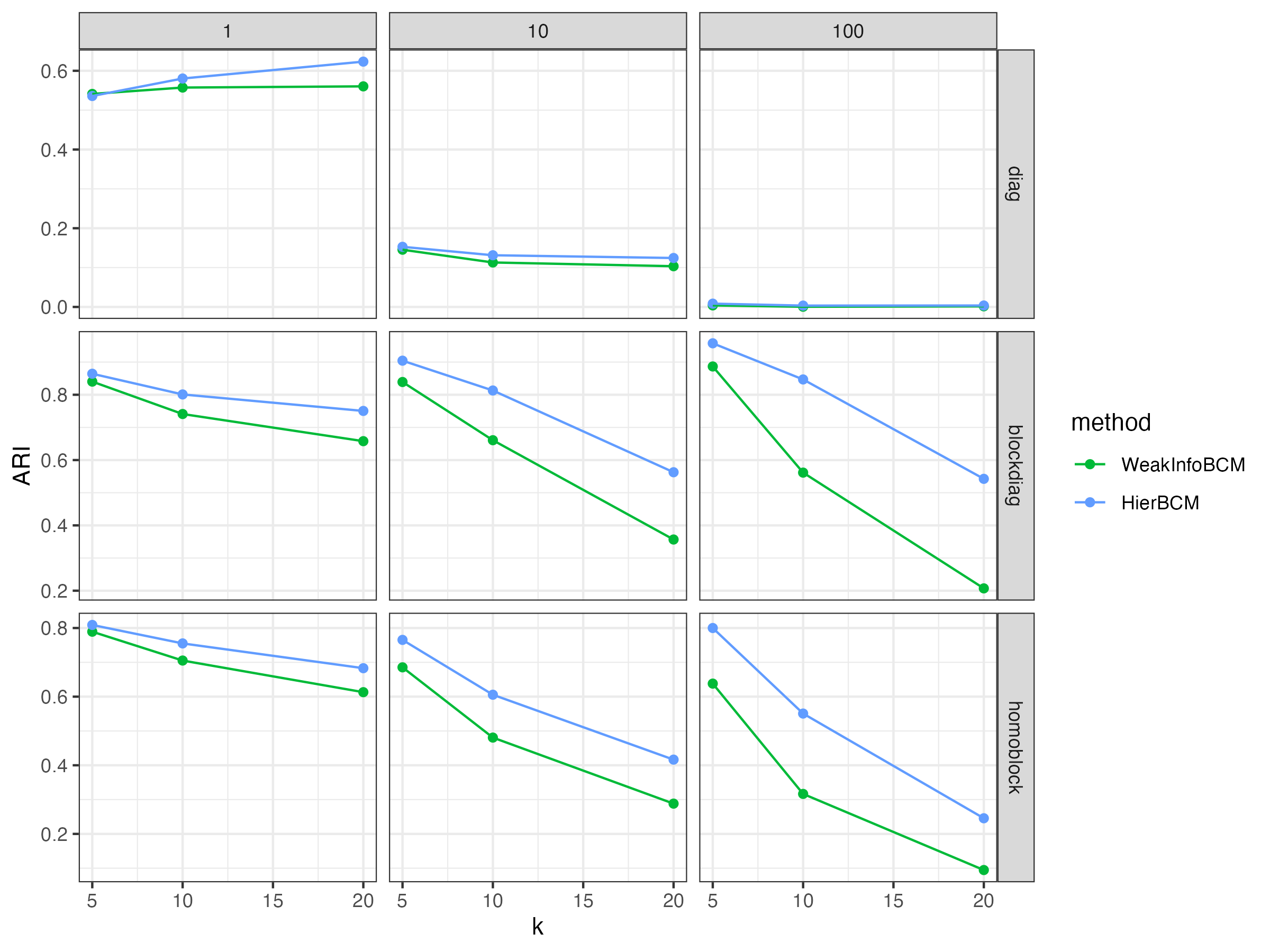}
\includegraphics[scale=.43]{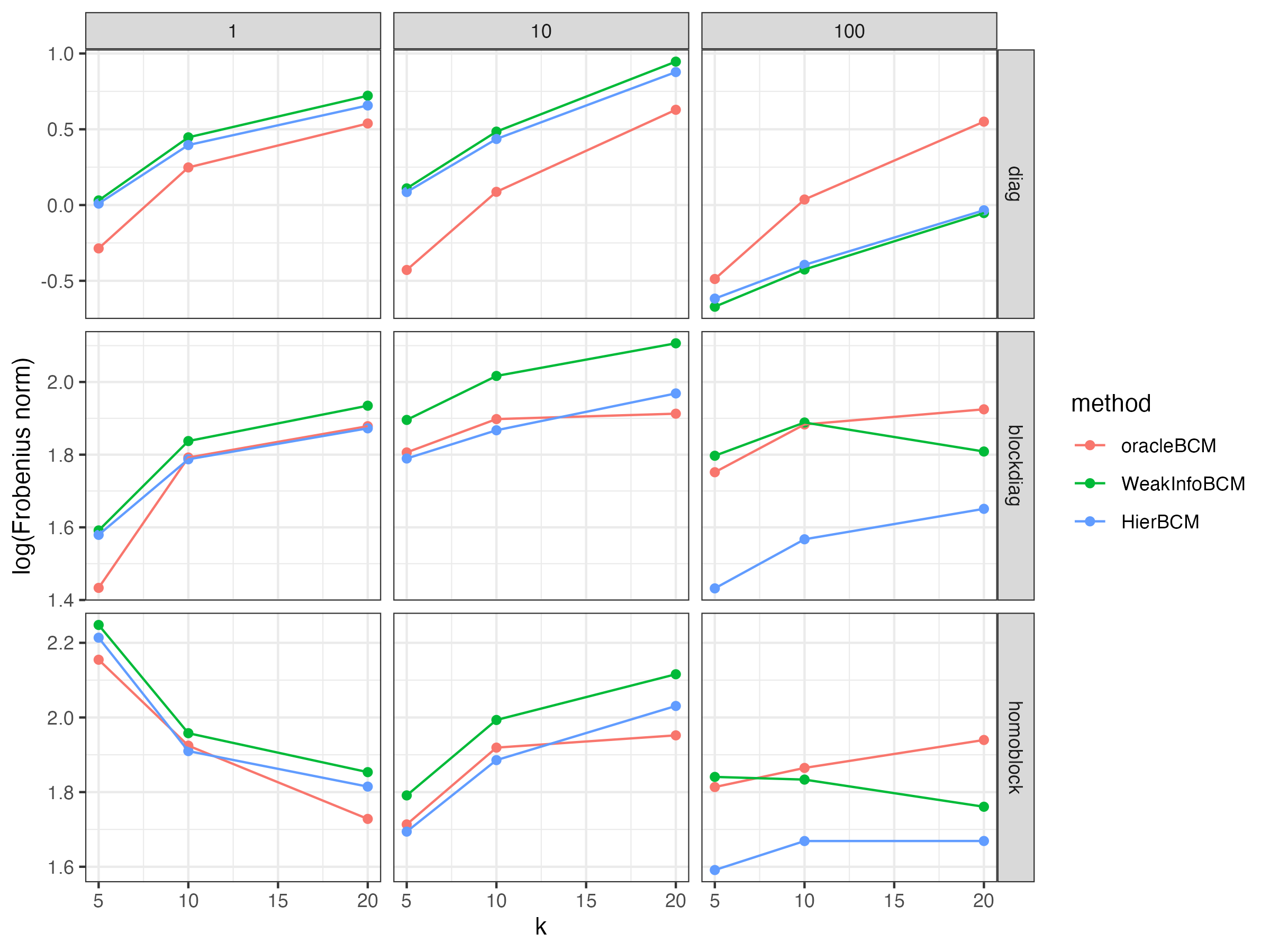}   
    
\end{center}
\caption[Estimation accuracy of different priors in the well-specified cases ($p=25$).]{Estimation accuracy of different priors in the well-specified cases, measured by ARI (left) and the logarithm of Frobenius distance. Each column represents a level of precision, indicating how closely the true covariance matrix aligns with the structures shown in each row. The hierarchical prior consistently outperformed or matched the weakly-informative prior across all scenarios, benefiting from its adaptive mechanism and better clustering performance. }
\label{fig:well.n25}
\end{figure}

Figure \ref{fig:well.n25} provides a visual summary of the performance of block covariance estimation under the weakly-informative prior and the hierarchical prior across all designs with $p=50$ and $n=25$. Two metrics were used to measure performance. The first measure was simply the Frobenius distance of $\|\hat\Sigma - \Sigma\|_F$, 
%
%
where $\|A\|_F = (\sum_{i,j=1}^p|a_{ij}|^2)^{1/2}$. 
The second measure evaluated accuracy of group allocation quantified as the Adjusted Rand Index (ARI) which measures similarity between two partitions by calculating the proportion of pairs of items that are either grouped together or separated in both partitions. 
%
%
\begin{equation}
    \text{ARI} = \frac{(p^2-p)\sum_{i,j}
        (m_{ij}^2-m_{ij})-\sum_i 
            (m_{i\cdot}^2-m_{i\cdot})\sum_j 
                (m_{\cdot j}^2-m_{\cdot j})}{(p^2-p) \{\sum_i (m^2_{i\cdot}-m_{i\cdot}) + (m_{\cdot j}^2-m_{\cdot j})\}/2 - \sum_i 
            (m_{i\cdot}^2-m_{i\cdot}) \sum_j 
                (m_{\cdot j}^2-m_{\cdot j})}
\end{equation}
where $i = 1,\cdots,r$, and $j = 1,\cdots,s$. Here, $m_{ij}$ represents the number of common elements between group $i$ in partition $\mathcal{B}$ and group $j$ in partition $\mathcal{B^*}$, and $m_i. = \sum_j m_{ij}$ and $m._j = \sum_i m_{ij}$. ARI ranges from -1 to 1, where a value of 1 indicates perfect alignment between the two partitions, while 0 suggests complete disagreement. ARI is invariant to group labels and remains robust regardless of the number of groups. 

Figure \ref{fig:well.n25} shows that the hierarchical prior consistently outperformed the weakly-informative prior both in terms of estimatio of $\Sigma$ and recovery of the block structure $\sc B$. Interestingly, it often approached or exceeded the performance of an oracle estimator of $\Sigma$ which assumed knowledge of the true partition $\sc B$. Clearly, relative to the weakly-informative prior, the hierarchical prior was more adaptive across various block structures. When the true $\Sigma$ resembled a diagonal matrix, serendipitously matching the prior mean of the weakly-informative prior in \eqref{prior:wi}, the two methods performed similarly. But in all other cases, the hierarchical prior performed better, potentially due to its ability to borrow information across blocks to offer better estimates of $\Sigma$ as well as a better scoring of $\sc B$ through $p(Y | \sc B)$. 
The larger the value of $\tau$, the more pronounced was this effect. 

\subsection{Broader performance analysis}

\subsubsection{Experimental design}
\label{sec:dgp}

In our second experiment we pitted the block covariance estimation method against a variety of covariance estimators widely used in the literature, allowing for the true covariance matrix $\Sigma$ to deviate from the grouped covariance assumption. Again, we generated data according to the Gaussian model $\bm y_i \sim N(0,\Sigma)$ under various assumptions on $\Sigma$ commonly found in the literature. The true $\Sigma$ remained fixed across 100 replicates for each scenario, which are shown in Figure \ref{fig:ex_dgp} corresponding to the sparsity assumptions listed below:
%
%
%
\begin{figure}[!t]
\begin{center}
\includegraphics[width=\linewidth]{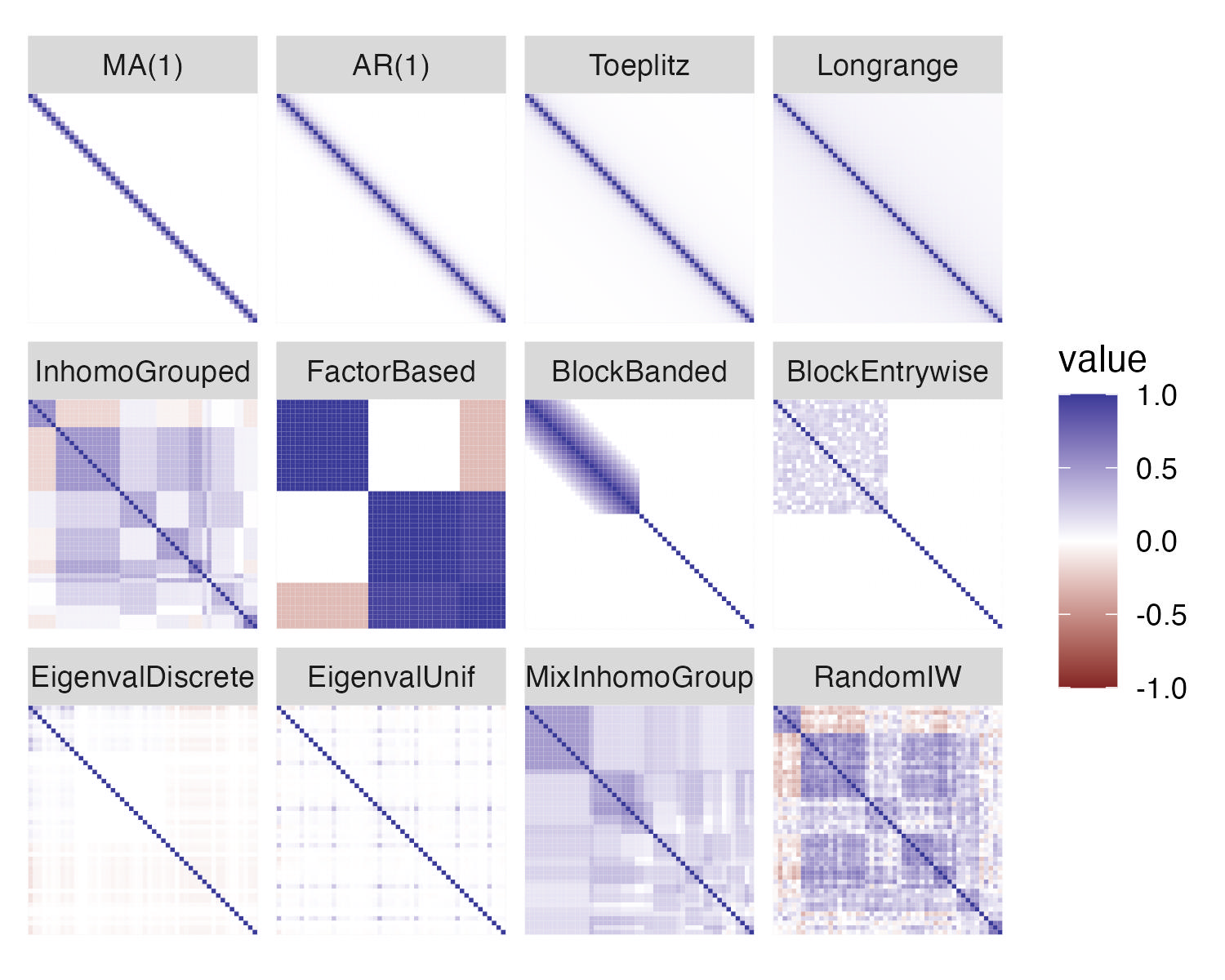} 
\end{center}
\caption[Example correlation matrices ($p=50$) generated for each data generating process.]{Example correlation matrices ($p=50$) generated for each data generating process. Each panel represents a distinct scenario, organized by the category of underlying assumptions. }
\label{fig:ex_dgp}
\end{figure}
\begin{enumerate}
    \item {\bf Order-dependent sparsity}, following \citet{bickel2008regularized} and \citet{cai2010optimal}, where the covariance $\sigma_{ij}$ decays as the distance between $i$ and $j$ increases. We examined four structures -- MA(1), AR(1), Toeplitz, long-range dependence matrices -- representing decay rates from rapid to slow. The parameters were set to $\rho = 0.5$, $H=0.7$, and $\alpha = 0.3$.

\begin{align}
\mbox{MA(1):}&~\sigma_{ij} =\rho^{|i-j|}\mathbf{1}\{|i-j|\leq 1\},\\
\mbox{AR(1):}&~\sigma_{ij} =\rho^{|i-j|},\\
\mbox{Long-range:}&~\sigma_{ij} =\frac{1}{2}[(|i-j|+1)^{2H}-2|i-j|^{2H}+(|i-j|-1)^{2H}],\\
\mbox{Toeplitz:}&~\sigma_{ij} =\rho|i-j|^{-(\alpha+1)}.
 \end{align}

\item{\bf Inhomogeneous grouped covariance matrix}, selected from a specific scenario of the grouped covariance matrix generating process described in \ref{sec:numerical_prior}. Specifically, we set $k^* = p/5$, $\tau = 10$, $(\delta_1,\delta_2,\delta_3)=(0.5,0.2,0.3)$. This inhomogeneous grouped covariance matrix was also used as the mean parameter in the inverse-Wishart distribution to generate random covariance matrices in other scenario.

\item {\bf Factor-based sparsity}, following \citet{zou2006sparse}, where data was generated from three latent factors: $f_1 \sim N(0,290), f_2 \sim N(0,300), f_3 = -0.3f_1 + 0.925 f_2 + \gamma, \gamma \sim N(0,1)$, where $f_1,f_2$, and $\gamma$ were independent. Observations were generated as $Y_i = f_j + \epsilon$, with $\epsilon \sim N(0,1)$, and the proportion of $Y_i$ drawn from $f_1$, $f_2$, and $f_3$ were set to 40\%, 40\%, and 20\%, respectively.

\item {\bf Block-sparsity}, following \citet{cai2012adaptive}, where the covariance matrix $\Sigma$ was block-diagonal: $\Sigma = diag(A_1,A_2)$, with $A_1$ and $A_2$ being $p/2$ by $p/2$ matrices. The first block, $A_1$, was diagonal with all entries set to 4. The second block, $A_2$, was considered under two structures: a banded matrix with entries $\sigma_{ij}=(1-|i-j|/10)_+$; entrywise-independent sparsity, where $A_2 = B + \epsilon I$, with $b_{ij}$ being independently drawn from $\text{Unif}(0.3,0.8)\times \text{Bernoulli}(0.2)$, and $\epsilon = \max(\lambda_{\min}(B),0)+0.01$ ensuring positive definiteness.

\end{enumerate}

Apart from these sparsity structures, we also considered lower-dimensional structures, following \citet{ledoit2020analytical}, where covariance matrices were generated from eigen decomposition. Eigenvalues were either taken to be discrete (40\%, 40\%, and 20\% of the eigenvalues were set to 10, 3, and 1, respectively), or drawn uniformly distributed between 1 and 10. Eigenvectors were randomly generated to form orthonormal matrices. Discrete eigenvalues tended to induce a block structure in the covariance matrices.

We also considered two other structures where the group covariance assumption was violated: (1) mixtures of block covariance matrices, where true covariance matrix was a mixture of three independently generated inhomogeneous grouped covariance matrices \(\textstyle (k^*,\tau,\delta_1,\delta_2,\delta_3)=(p/5,10,0.5, \allowbreak 0.2,0.3)\) with equal mixing weights. (2) random Inverse-Wishart distributed covariance matrices, where the true covariance matrix was drawn from $IW(\nu_0,(\nu_0-p-1) \Sigma_{ig})$, centered at the previously described inhomogeneous grouped covariance matrix $\Sigma_{ig}$. The precision parameter $\nu_0$ controlled deviation from the group structure assumption, and we set $\nu_0 = p+2$.

To facilitate comparison among different estimators and data generating processes, we used the sample covariance matrix as a benchmark, assessing performance by the ratio of the Frobenius norm of each estimator to that of the sample covariance matrix:
\begin{equation}
    \text{R}_{est/sample} = \frac{ \|\hat{\Sigma} - \Sigma\|_F}{\|S - \Sigma\|_F}.
    \label{equ:rfn}
\end{equation}
%
A lower ratio indicates better estimator performance, with a ratio of 1 suggesting similar performance to the sample covariance matrix.

\subsubsection{Competing estimators}

We considered block covariance matrix estimators under three priors: weakly-informative, MLE-equivalent, and our proposed hierarchical prior. For each method, estimates were obtained with 5000 Markov chain Monte Carlo iterations, discarding the first 500 as burn-in, and every fifth posterior sample retained. The posterior mean served as the estimator for Bayesian methods. 

For alternative covariance estimators, we utilized methods available in the R package \texttt{cvCovEst} \citep{boileau2022cvcovest}, which can be categorized based on their underlying assumptions:

\begin{itemize}
    \item order-dependent sparsity: Banding \citep{bickel2008regularized} and tapering estimators \citep{cai2010optimal} assumes that covariance decays with distance between variable indexes.
    \item order-invariant sparsity: hard thresholding \citep{bickel2008regularized}, SCAD thresholding, and adaptive LASSO estimators \citep{zou2006adaptive, rothman2009generalized} assume sparsity without dependence on variable ordering.
    \item shrinkage estimators: Ledoit-Wolf linear and nonlinear shrinkage estimators do not impose structural assumptions but shrink sample eigenvalues toward a target. \citep{ledoit2004well,ledoit2020analytical}.
\end{itemize}
The banding estimator applies hard thresholding to off-diagonal entries of the sample covariance matrix based on the distance between the positions of the variables within a user-specified ordering, setting distant covariances to zero. These are most appropriate for time series data where a linear ordering is known and is taken to provide information about relative strength of correlation between variables. The tapering estimator is similar to the banding estimator but instead of hard thresholding, it gradually shrinks off-diagonal entries with increasing distance, resulting in a dense matrix with diminishing covariances. Hard thresholding and SCAD estimators apply thresholds to individual entries of the sample covariance matrix -- no particular order of presentation is assumed. The adaptive LASSO estimator penalizes entries unequally using data-dependent weights. Ledoit-Wolf linear shrinkage applies uniform shrinkage to all sample eigenvalues, pulling them toward a central value, while the nonlinear shrinkage estimator adjusts the shrinkage intensity based on the distribution of the sample eigenvalues, allowing for individualized shrinkage.

\subsubsection{Comparison of estimators across various covariance structures}
\label{sec:numerical_est}

\begin{figure}[ht]
\begin{center}
\includegraphics[width=0.9\linewidth]{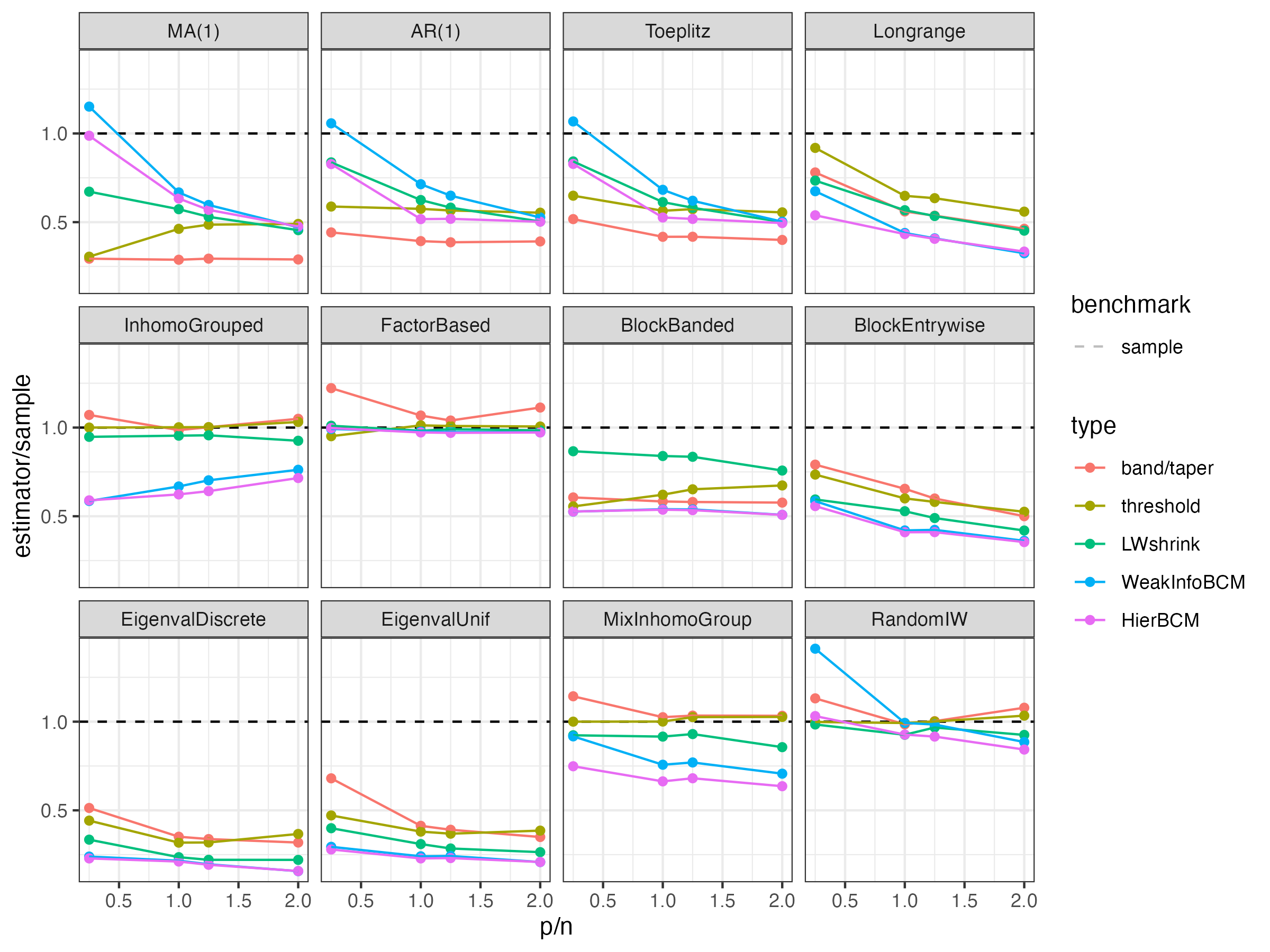}  
\end{center}
\caption[Comparison of estimation accuracy of different covariance estimators across various covariance structure ($p=50$)]{Comparison of estimation accuracy among different covariance estimators across various covariance structures ($p=50$). The block covariance estimator with the hierarchical prior (lavender line) generally outperformed alternative estimators, except in cases with order-dependent structures featuring rapidly decaying dependencies. This result underscored the adaptability of the block structure assumption across diverse covariance types. In addition, the hierarchical prior consistently surpassed both the weakly informative prior and the sample covariance matrix.}
\label{fig:rfn.p50}
\end{figure}

We examined the performance of different estimators under various covariance structures using $p = {20,50}$\footnote[1]{See Figure \ref{fig:rfn.p20} for case $p=20$ in the Appendix.}, and $p/n = {0.1, 1, 1.25, 2}$. For clear comparison, we grouped alternative estimators by underlying assumptions and selected the best-performing one from each category. Results are presented in Figure \ref{fig:rfn.p50}. It can be seen that the block covariance estimator with the hierarchical prior (lavender line) and weakly-informative prior (blue line) generally outperformed alternative estimators, except in cases involving order-dependent structures with rapidly decaying dependencies and randomly generated Inverse-Wishart covariances. This result highlights the adaptability of the group covariance matrix assumption across diverse covariance types. Notably, the hierarchical prior consistently surpassed the weakly-informative prior, demonstrating its capacity to flexibly accommodate different structural patterns in the data.

For covariance with order-dependent sparsity, the banding/tapering estimator consistently performed better than other estimators, except in cases with long-range dependencies. This is because the banding/tapering estimators correctly bet on the underlying structure, whereas other estimators struggled due to misspecification. Moving from left to right across the first row of panels, as dependency decayed more slowly, covariance matrices had less sparsity and larger off-diagonal values (as shown in Figure \ref{fig:ex_dgp}). In such cases, the assumptions underlying the banding/tapering and thresholding estimators became less appropriate, the block covariance estimator performed better, particularly as long-range dependencies grew stronger.

For random Inverse-Wishart distributed covariance matrices, our estimator with a hierarchical prior performed comparable to, or outperformed, the best alternative (namely, the Ledoit-Wolf shrinkage estimator in cases where $p/n=0.1$). In contrast, the weakly-informative prior often performed worse than, or on par with, the sample covariance when $p \leq n$. This might be attributed to the lack of clear structure in randomly generated matrices (see Figure \ref{fig:ex_dgp}), which deviated from group covariance matrix and sparsity assumptions. Unlike the hierarchical prior, the weakly-informative prior lacked adaptivity, making it less robust against misspecification.

Across all scenarios, our estimator outperformed the sample covariance matrix. Notably, the weakly-informative prior sometimes underperformed relative to the sample covariance matrix, particularly when $n > p$ and for Inverse-Wishart or rapidly decaying, order-dependent structures. This result illustrates the limitations of a fixed shrinkage target and the necessity of adjustable shrinkage intensity. The hierarchical prior appears to provide adaptability in shrinkage, ensuring robust performance even under challenging conditions.

\section{Relation to Other Covariance Estimation Methods}

At this point, a comprehensive mathematical explanation remains elusive as to why the grouped covariance assumption works so well across various challenging scenarios. Here we shed further light on the the underlying theory by drawing attention to other theoretical frameworks commonly adopted in the literature which seem to have close relations to the stochastic block covariance model.


\subsection{Clustered factor model}
\label{sec:cfm}
Stochastic block covariance estimation shares close relations to regularized factor models where additional structure is added by enforcing identical factor loadings across variables. This idea has been explored in the literature under the name of {\it clustered factor models} \citep{taylor2017joint, tong2023characterizing}. Suppose $\bm{y}_i \overset{\text{i.i.d}}{\sim} N(\bm{0},\Sigma)$, where $\Sigma$ is a group covariance matrix satisfying \eqref{equ:model}. The data can be modeled using a latent factor model:
\begin{equation}
    \bm{y}_i = L \bm{f} + \bm{e}_i,
\end{equation}
where $\bm{y}_i$ is a $p \times 1$ vector, $L$ is a $p\times r$ matrix of factor loadings, $\bm{f}$ is a $r \times 1$ vector of latent factors with $E(\bm{f})=\bm{0}$, and $\var(\bm{f})=I_r$, and $\bm{e}_i\overset{\text{i.i.d}}{\sim} N(\bm{0},E)$ is the error term with $E$ being a diagonal matrix. The covariance matrix can then be expressed as 
\begin{equation}
    \Sigma = LL^T+E,
\end{equation}
with factor loadings identical for variables within the same block. 

In spite of the similarity to stochastic block covariance model, as an estimation framework, the clustered factor model is more restrictive and less efficient in its representation. First, it represents a subset of stochastic block covariance matrices that are positive definite (see Theorem 1 and Lemma 1 in \citet{tong2023characterizing}), and usually requires more parameters ($k\times r + r$) than stochastic block covariance matrices, except in cases where the latter is highly parsimonious ($k < 2r - 3$). Additional challenges present in deciding on an optimal number of factors and in interpreting the estimated factors and their loadings. 
%
Stochastic block covariance representation, in contrast, provides a more straightforward estimation and transparent interpretation.

\subsection{Stein's class of shrinkage estimators}

In our experiments reported above, the Ledoit-Wolf shrinkage estimator consistently ranked as the second-best method, following closely behind the block covariance matrix estimator in most cases, except for those involving banded/tapered or sparse banded structures. It outperformed or remained competitive with the block covariance estimators when the true covariance matrix was generated from an inverse-Wishart distribution or followed an MA(1) or AR(1) process in settings where $n > p$. Notably, it consistently surpassed the sample covariance matrix even in challenging cases. We offer here a preliminary analysis of the similarities and differences between the Ledoit-Wolf estimator and our stochastic block covariance estimator. 

The Stein's class of shrinkage estimators, originally introduced by \citet{Stein1975, Stein1986}, shrinks the sample eigenvalues while retaining the sample eigenvectors. The rationale behind this approach is that sample eigenvalues tend to be more dispersed than population eigenvalues, and appropriate shrinkage can improve estimation accuracy. Let $S$ denote the sample covariance matrix with eigen decomposition $S=U\Lambda U'$. Stein's shrinkage estimators take the form
\begin{equation}
    \hat{\Sigma}=U \Delta U',
\end{equation}
where $U$ represent the eigenvector of $S$, and $\Delta = \text{diag}(\delta_1,\dots,\delta_p)$ consists of modified sample eigenvalues, where each $\delta_j$ is a function of the sample eigenvalues $\Lambda$, i.e., $\delta_j = \varphi_j(\Lambda)$, mapping from the positive diagonal matrix to itself.  


We note that the block covariance estimator shares some conceptual similarities with Stein's estimators, since both reduce the dispersion of sample eigenvalues, albeit through different mechanisms. While Stein's approach shrinks eigenvalues directly, guided by the rotation of the sample eigenvectors, the block covariance estimator achieves shrinkage by clustering eigenvalues, driven by a block-sparse rotation, effectively eliminating within-group variance. 

We compared the performance of block covariance estimator against various shrinkage estimators within Stein's family, including the population eigenvalues plug-in estimator ($\Delta = \Lambda$), Finite Sample OPTimal estimator \citep{ledoit2012nonlinear}($\delta_j = u_{j}' \Sigma u_{j}$), and the analytical nonliear Ledoit-Wolf estimator \citep{ledoit2020analytical}. Our numerical experiments focused on cases where $p > n$, with $p/n = 1.1, 2, 5, 10$. We followed the same data generating processes described in Section \ref{sec:numerical_est}, except that the factor-based sparsity scenario was replaced with a degenerate covariance matrix where all eigenvalues were equal to 1. 

\begin{figure}[ht]
\begin{center}
\includegraphics[width=0.9\linewidth]{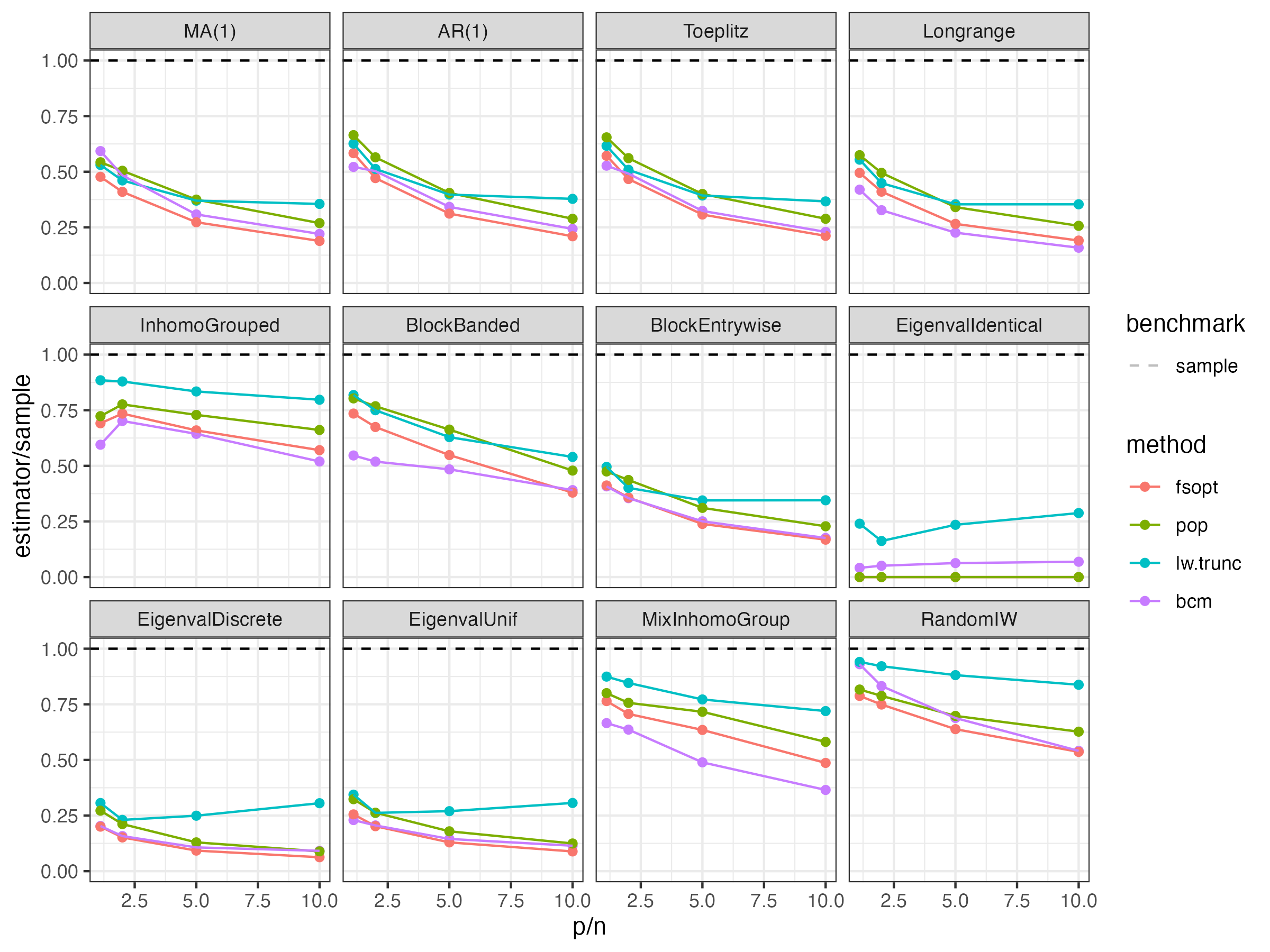}  
\end{center}
\caption[Comparison of estimation accuracy of block covariance estimation with Stein's class of shrinkage estimators across various covariance structure ($p=50$)]{Comparison of estimation accuracy of block covariance estimation with Stein's class of shrinkage estimators across various covariance structure ($p=50$). The block covariance estimator generally outperformed or was comparable to oracle estimators, except in cases where the true covariance matrix followed an MA(1) process or a random inverse-Wishart distribution.}
\label{fig:rfn.stein.p50}
\end{figure}

Figure \ref{fig:rfn.stein.p50} displays a comparison of estimation accuracy between stochastic block covariance estimator (BCM) and Stein's class of estimators across various covariance structures. \footnote{To prevent numerical instability in density estimation, we truncated sample eigenvalues below $10^{-8}$ to zero when applying the Ledoit-Wolf method. } The block covariance estimators generally outperformed or was comparable to the oracle estimators, except when the covariance matrix was generated from an MA(1) process (characterized by rapid decay in dependency) or a random inverse-Wishart distribution. Across nearly all scenarios, the stochastic block covariance estimator consistently surpassed Ledoit-Wolf estimator. This advantage arose from the BCM’s reliance on rotation learning, which effectively leveraged structural information within the data, allowing better adaptation when such structure was present. In contrast, Stein’s family estimators operated without strong structural assumptions, making them overly conservative and limiting their flexibility and adaptability to various covariance matrix configurations.

The stochastic block covariance estimator and Stein's estimator originate from two classical models: factor analysis and principle component analysis (PCA), respectively. Our model introduces additional regularization on factor loadings by enforcing identical factor loadings within the same cluster while also promoting sparsity. Similar extensions of factor analysis and PCA can be found in \cite{rohe2023vintage}, where varimax rotation is recommended to promote sparsity and interpretability for non-Gaussian data.

\section{Application} 

\subsection{Neural coordination}
\label{sec:neuro}

Previous studies \citep{caruso2018single, jun2022coordinated, schmehl2024multiple, chen2024spike} have shown that individual neurons can encode two simultaneously presented stimuli through a turn-taking dynamics called "code juggling", where neurons stochastically switch responding to two composite stimuli, such as at either a slow rate across trials or a fast rate within a trial. However, this alternation between stimuli is not entirely random. Some evidence suggests a coordinated pattern of code juggling across neural populations \citep{jun2022coordinated, schmehl2024multiple}. This coordination can be represented by a block structure in covariance matrices, which may provide valuable insights into how neural circuits contribute to perception and behavior, a fundamental question in neuroscience.

\begin{figure}[!ht]
\begin{center}
\includegraphics[width=\linewidth]{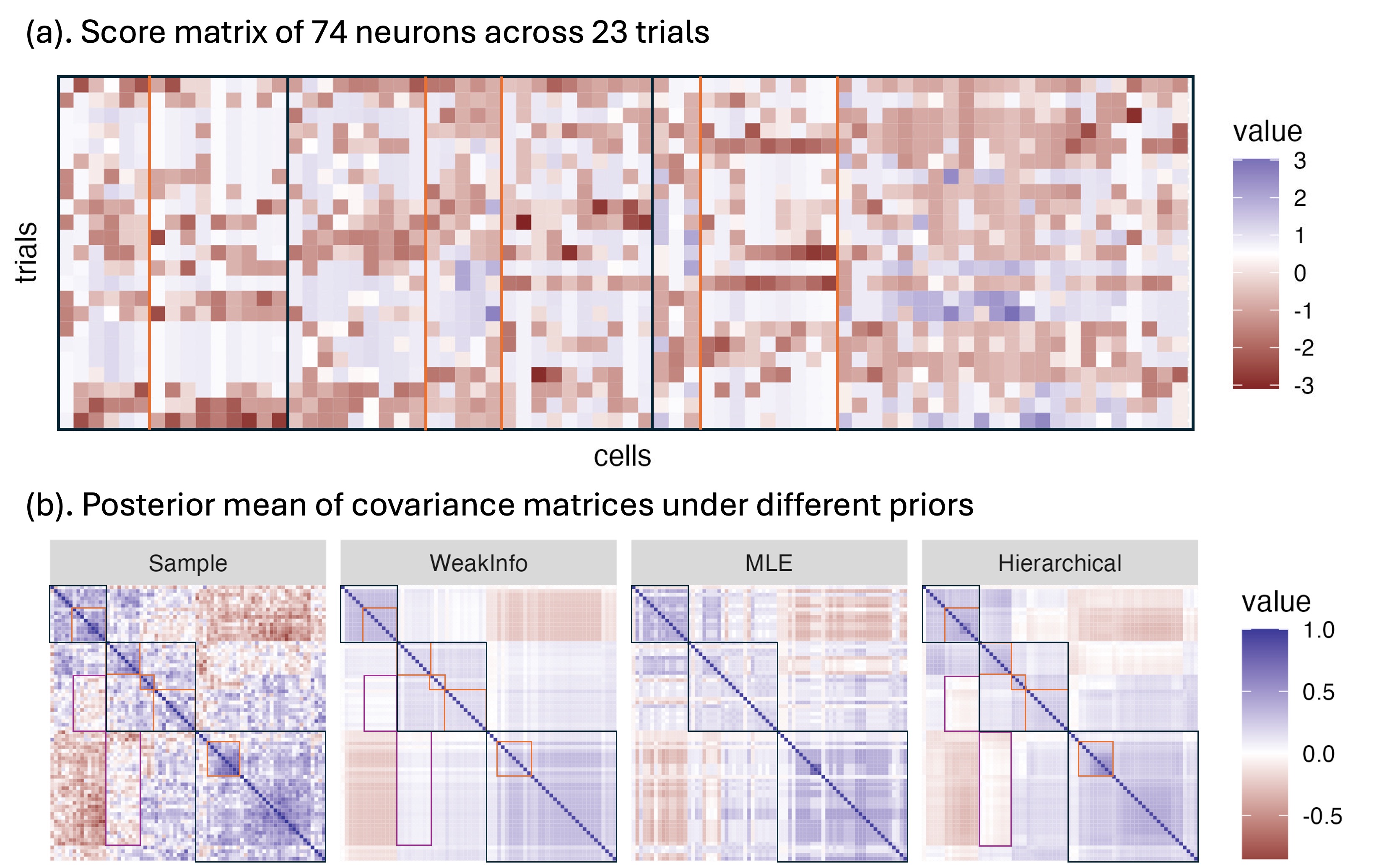}    
\end{center}
\caption[(a)Score matrix of 74 neurons across 23 trials. (b) Posterior mean of covariance matrices under different priors.]{(a) Score matrix of 74 neurons across 23 trials, with each entry showing the neuron’s tendency to encode stimulus A (purple) or B (red) during simultaneous dual-stimulus exposure. Co-fluctuating and oppositely fluctuating patterns among neurons suggested clustering in the covariance structure. (b) Posterior mean of covariance matrices under different priors. Neurons in (a) and (b) were sorted in the same order based on clustering results from stochastic block covariance estimation under the weakly-informative prior and the hierarchical prior. Black boxes outline the three main groups (size 15, 24, and 35) clustered under the weakly-informative prior. Within these black boxes, orange boxes represent sub-group structure detected by the hierarchical prior but ignored by the weakly-informative prior. The weakly informative prior oversimplified the block structure (orange) and misrepresented between-block covariances (purple), while the MLE-equivalent prior favored sparsity, resulting in many isolated small-sized blocks shown as white ribbons. The hierarchical prior, with adaptive shrinkage, refined the block structure to better reflect the underlying data. }
\label{fig:neuron_all}
\end{figure}


We analyzed a dataset from \citet{ruff2016attention}, where monkeys performed a motion direction change detection task while neural activities were recorded. In trials involving dual stimuli (both A and B), adjacent drifting Gabor patches with different orientations were presented. In single stimulus trials, only one of the constituent drifting Gabor patches was visible. A multi-electrode array was planted in V1 area recording spike activities of 101 neurons simultaneously. We used a subset of the data and preprocessed it to generate a score matrix for 74 neurons across 23 trials, where each entry reflects a neuron's inclination to encode stimulus A. For a detailed description of the preprocessing procedure, refer to the Appendix \ref{append:neuro}.

Figure \ref{fig:neuron_all}(a) displays the score matrix, revealing distinct clusters of neurons with different patterns in their activity fluctuations. Examining the underlying block structure of these fluctuations was our primary focus. Figure \ref{fig:neuron_all}(b) shows posterior mean of covariance matrices comparing shrinkage effects under different choices of hyperparameters. Figure \ref{fig:neuron_psm} in the Appendix presents the posterior similarity matrix, which quantifies the posterior probability of each neuron pair being grouped together. Neurons in these three figures were arranged in the same order. Black boxes outline the three main groups identified under the weakly-informative prior. Within these groups, orange boxes highlight detailed sub-groups captured by the hierarchical prior but overlooked by the weakly-informative prior. Purple boxes highlight covariances that were misrepresented by the weakly-informative prior but well captured by the hierarchical prior. 

To facilitate interpretation, special care is required for neuron reordering. For better comparison among different choices of hyperparameters, we incorporated clustering information from both the weakly-informative prior and the hierarchical prior using a two-stage procedure. We first reordered neurons based on hierarchical clustering, where the distance matrix was defined as 1 minus the mean of posterior covariance matrices under the weakly-informative prior. In this process, three major blocks emerged, outlined by black boxes in Figure \ref{fig:neuron_all}(b). We then repeated the same procedure to reorder neurons within each block, but with respect to mean of posterior covariance matrices under the hierarchical prior instead of the weakly-informative prior. A finer structure ignored by the weakly-informative prior was revealed by the hierarchical prior, highlighted by orange and purple boxes in Figure \ref{fig:neuron_all}(b).


In Figure \ref{fig:neuron_all}(b), three primary clusters emerged: a median-sized cluster of 15 neurons and two larger clusters of around 24 and 35 neurons, as outlined by black boxes. Neurons within the first cluster were highly correlated, positively correlated with the second cluster and negatively correlated with the third.

The weakly informative prior excessively shrank the covariance toward a block diagonal structure, oversimplifying the block structure and misrepresenting between-block covariances (particularly negative ones). As shown in Figure \ref{fig:neuron_all}(b), the weakly informative prior manifested two main issues in covariance estimation: (1) diminished within-block covariances, where some high positive within-block covariances among certain neurons were significantly reduced due to the shrinkage (see orange boxes). (2) a reversal of the sign of between-block covariances, where some negative covariances were shrunk to weak positive covariances (see purple boxes). The large bias in covariance estimation was related to the oversimplified block structure estimation (see Figure \ref{fig:neuron_psm}). The weakly informative prior underestimate of the block count $k$, weakening the effective use of structural information on covariance estimation.

The MLE-equivalent prior focused on capturing the sparsity of the sample covariance matrix, resulting in the formation of too many blocks, particularly isolated small-sized ones. In Figure \ref{fig:neuron_all}(b), the numerous white ribbons in the covariance estimator indicated that many entries were overly shrunk toward zero, especially within small blocks. Comparing the posterior similarity matrix in Figure \ref{fig:neuron_psm} with the sample covariance matrix in Figure \ref{fig:neuron_all}(b), it appeared that the MLE-equivalent prior prioritized sparsity over preserving the block structure. The block assignment was largely driven by its sparsity, rather than by meaningful block patterns. Additionally, the posterior similarity matrix of the MLE-equivalent prior showed a concentrated distribution, which may suggest slow MCMC mixing or a tendency to get stuck in a local mode.

Both the weakly informative and MLE-equivalent priors suffered from the mis-specification of the shrinkage target and lack adaptive intensity. In contrast, the hierarchical prior refined the block structure to better align with the data through adaptive shrinkage (see Figure \ref{fig:neuron_all}). Unlike the weakly informative prior, the hierarchical prior was sensitive to the heterogeneity of between-block covariances (purple boxes), and effectively captures finer sub-block structures (orange boxes). Unlike the MLE-equivalent prior, the hierarchical prior avoided introducing isolated small blocks and excessive shrinkage toward zero, providing a smoother, more accurate estimator. Figure \ref{fig:neuron_psm} further illustrated how the hierarchical prior enhanced cluster segmentation and restored essential features. This refined block structure affirmed the presence of coordination within the neural population, providing a clear depiction of potential neural circuits.

\subsection{Monthly stock returns of energy and financial companies}
\label{sec:enron}
Estimating the covariance matrix is crucial in the financial market, particularly for portfolio optimization and risk management. An accurate covariance-based classification can provide insights into the interconnections among different assets, facilitating portfolio diversification to mitigate specific risks.

To illustrate covariance-based classification, we revisited the example from \citet{liechty2004bayesian} but addressed in a flexible way without prefixing the number of groups. In this example, we examined monthly stock returns spanning April 1996 to December 2001 for 9 companies (68 time points). These companies belonged to either energy sector (Reliant, Chevron, British Petroleum, and Exxon) or finance sector (Citi-Bank, Lehman Brothers, Merrill Lynch, and Bank of America). We only focused on data prior to Enron's bankruptcy (December 2001), differing from the analysis presented in \citet{liechty2004bayesian}.

\begin{figure}[!ht]
\begin{center}
\includegraphics[width=0.7\linewidth]{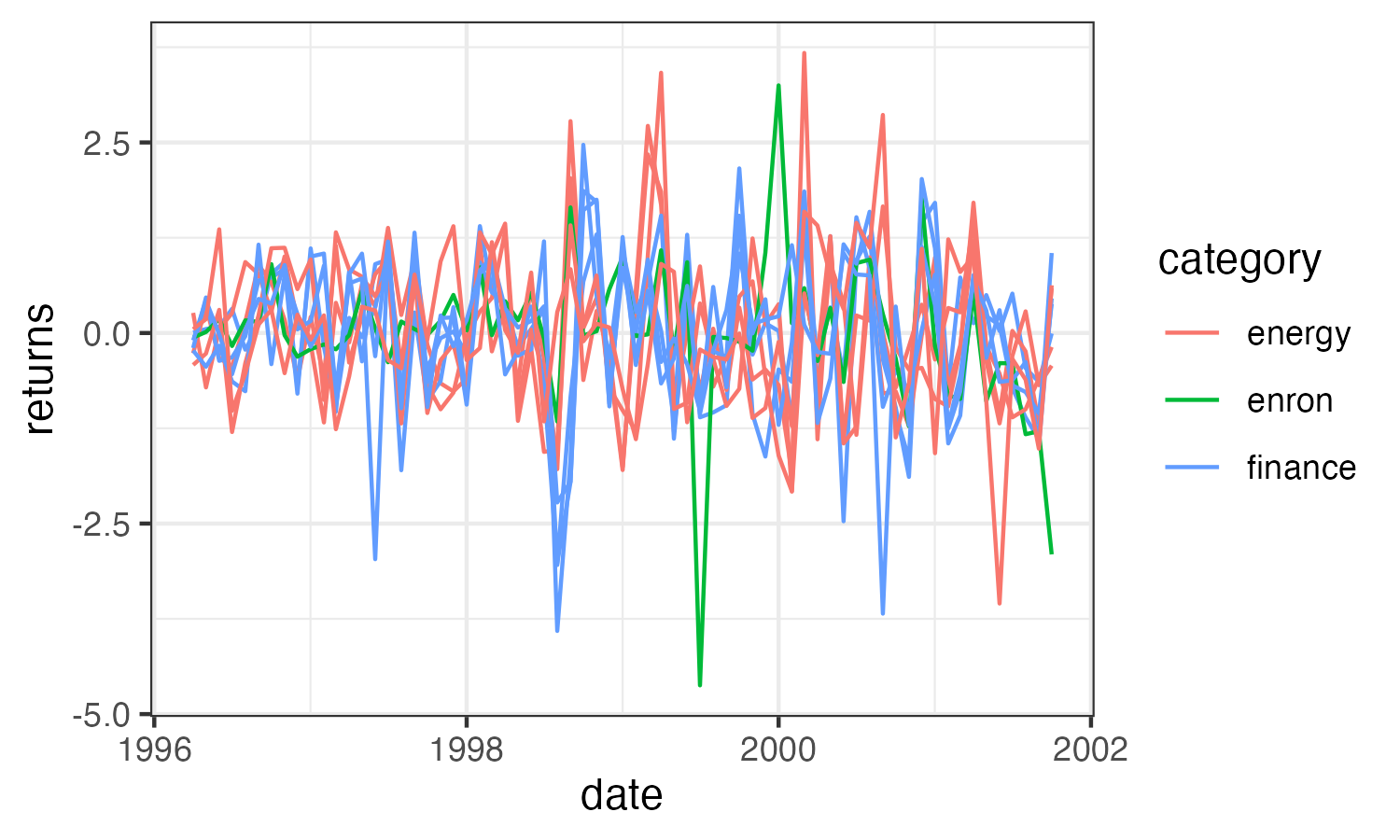}    
\end{center}
\caption{Monthly returns of Enron and other energy and finance companies. It was unclear whether Enron's fluctuations were more in sync with finance or energy companies}
\label{fig:enron_trace}
\end{figure}

We aimed to: (1) cluster these 9 companies to assess whether their variations align with their industrial classifications, and (2) evaluate Enron's transition from the energy sector to the finance sector prior to its bankruptcy. Given the significant fluctuations in some stocks, we assumed group covariance structure in the correlation matrix, applying our estimator to the standardized data. 

\begin{figure}[!ht]
\begin{center}
\includegraphics[width=\linewidth]{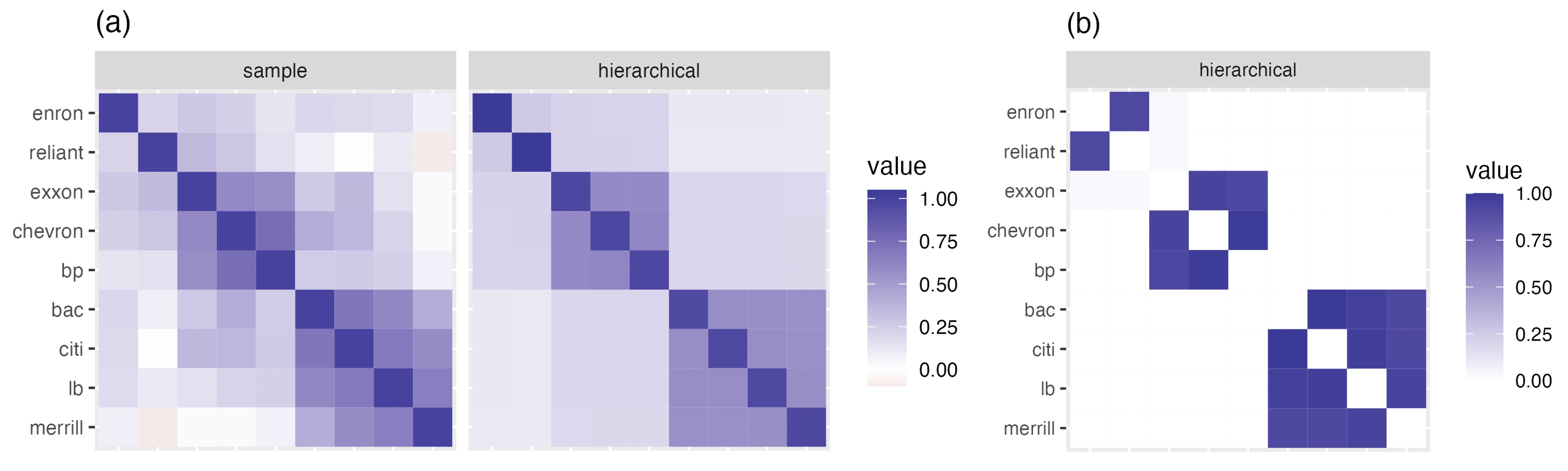}    
\end{center}
\caption[Estimators of covariance matrix and posterior similarity matrix.]{Estimators of covariance matrix (left: sample covariance, right: stochastic block covariance estimation under the hierarchical prior) and posterior similarity matrix. Enron behaved more like energy companies than finance companies.}
\label{fig:enron_all}
\end{figure}

Figure \ref{fig:enron_trace} displays a trace plot of standardized monthly returns, presenting a challenge to discern whether Enron's fluctuations were more in sync with finance or energy companies. The stochastic block covariance estimation, shown in Figure \ref{fig:enron_all}, revealed three main clusters: Enron and Reliant (energy), other energy companies, and finance companies. These classification aligned with their respective industrial classifications.

It appeared that Enron did not successfully transit from the energy to finance sector, as it behaved more closely resembled energy companies than finance companies. Enron and Reliant shared similarities, exhibiting lower covariance with other energy companies and finance companies compared to the remaining energy companies. This finding was consistent with the conclusion of \citet{liechty2004bayesian}, which also indicated Entron' unsuccessful transition.

\subsection{Phenotypic integration of Mediterranean shrubs under contrasting environmental conditions}
\label{sec:plant}

Analyses of correlation structures among traits can offer insights into phenotypic integration, i.e., {\it ``the pattern of functional, developmental, and/or genetic correlation among different traits in a given organism''} \citep{pigliucci2003phenotypic}. Traits may be positively or negatively correlated depending on their functional roles, influencing how organisms adapt to varying environmental conditions. To examine phenotypic integration under different environmental conditions, we present here an analysis of the correlation matrices of 20 ecophysiological traits in Mediterranean shrub samples (Lepidium subulatum L., Brassicaceae) grown under both stressful (drought) and favorable conditions with a focus on assessing the differences in phenotypic integration between these environments. The experimental details are described in \citet{matesanz2021phenotypic}. 

The dataset recorded traits of 355 individual plants across 4 populations, with around 15 to 20 families nested within each population. We aggregated plant-level data to the family-level by computing the average trait values for each maternal family within each population. After removing missing cases, we obtained 45 samples from plants grown under drought conditions and 55 from those in favorable conditions. To better meet the normality assumptions, we applied a square-root transformation to two traits (FN, PH), and a logarithmic transformation to eight traits (FBF, FL, IN, LA, LL, RO, TB, TELA) before standardization. These 20 traits could be categorized into the following functional groups: 
\begin{itemize}
    \item Morphological traits: specific leaf area (SLA), leaf area (LA), leaf length (LL), total estimated leaf area (TELA), plant height (PH)
    \item Physiological traits: midday photochemical efficiency of PSII (FvFm), relative growth rate (RGR)
    \item Phenological traits: onset of flower bud formation (FBF), onset of flowering (FL), onset of fruiting (FR), percentage of senescent leaves (Sen)
    \item Allocation traits: root-to-leaf ratio (R.L), leaf-to-stem ratio (L.S), above-ground biomass (AB), total biomass (TB)
    \item Reproductive traits: number of inflorescences (IN), inflorescence size (IS), number of flowers (FN), reproductive output (RO), seed size (SS)
\end{itemize}

These traits might either compete for resources or work together in a coordinated manner.

\begin{figure}[!ht]
\begin{center}
\includegraphics[width=0.8\linewidth]{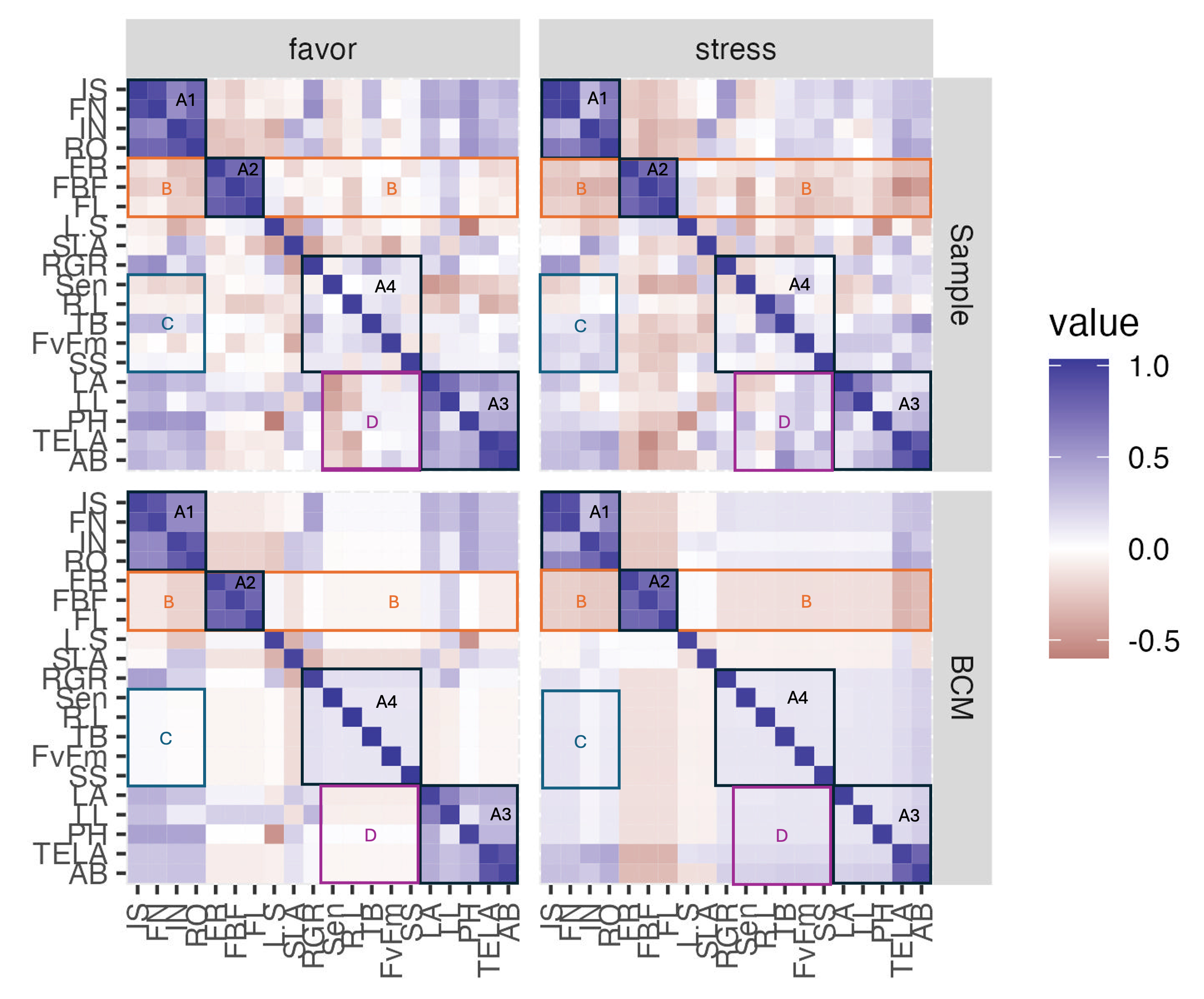}
\end{center}
\caption[Posterior mean of covariance matrices in the phenotypic integration application.]{Posterior mean of covariance matrices of traits under favorable and stressful environments. The overall block structure aligned well with functional group classifications, with notable differences in correlation patterns across environments (A1, A2, A3, A4). Under drought conditions, cross-functional correlations strengthened, suggesting an adaptive response to environmental stress (B, C, D).}
\label{fig:plant_cov}
\end{figure}

Figure \ref{fig:plant_cov} presents the sample correlation matrices and their posterior mean estimates under favorable and stressful conditions. The overall block structure aliged well with functional group classifications. Notably, reproductive traits (IS, FN, IN, RO) exhibited strong positive correlations, reflecting coordinated investment in reproduction (Figure \ref{fig:plant_cov} A1). Similarly, phenological traits (FR, FBF, FL) formed a highly correlated cluster (Figure \ref{fig:plant_cov} A2). These traits occurred in sequence during the reproductive timeline, which explained their strong correlation. The morphological traits (LA, LL, PH, TELA) were positively correlated with above-ground biomass (AB), indicating an integrated response to plant growth (Figure \ref{fig:plant_cov} A3). Additionally, a cross-functional cluster emerged, incorporating physiological (RGR, FvFm), allocation (R.L, TB), phenological (Sen), and reproductive (SS) traits (Figure \ref{fig:plant_cov} A4).

Under drought conditions, the correlation matrix exhibited enhanced correlation across functional groups compared to the favorable environment: (1) stronger negative correlations between reproductive phenology (FBF, FL, FR) and other traits (Figure \ref{fig:plant_cov} B); (2) greater positive correlations between reproductive traits (IS, FN, IN, RO) and traits in the cross-functional group (Sen, R.L, TB, FvFm, SS) (Figure \ref{fig:plant_cov} C); (3) enhanced positive correlations between above-ground growth traits (LA, LL, PH, TELA, AB) and traits in the cross-functional group (Sen, R.L, TB, FvFm, SS) (Figure \ref{fig:plant_cov} D). 

This pattern was consistent with a drought-escape strategy \citep{blanco2022natural, matesanz2020high}, wherein plants accelerate reproductive timing and enhance growth to complete life cycle before drought conditions become more severe. For example, a larger cross-functional cluster emerged, involving coordinated changes in physiological, phenological, and allocation traits. Under drought conditions, plants that began reproduction earlier tended to grow larger (higher PH, TB, LA), produced larger seeds, showed increased leaf senescence, and exhibited higher growth rates and photochemical efficiency. 

\citet{matesanz2021phenotypic} quantified phenotypic integration in plants by counting the number of significantly non-zero pairwise correlations between traits and found that plants under stressful conditions exhibited higher phenotypic integration. In our model, this increased integration appeared as a reduction in the number of distinct trait clusters (see posterior similarity matrix in Figure \ref{fig:plant_psm} for a clear block assignment). Under drought conditions, traits from different functional groups that were only weakly or negatively associated in favorable conditions became more strongly or positively associated, suggesting an adaptive reconfiguration of trait relationships to cope with environmental stress.

\section{Discussion}
\label{sec:discussion}

We introduced a stochastic block structure for covariance matrix estimation, motivated by practical demands in real-world applications. Beyond its interpretability, the stochastic block structure effectively reduces dimensionality, while being more general, flexible, and efficient compared to conventional sparsity assumptions or shrinkage estimators. The stochastic block structure also offers greater adaptability across covariance matrices with different structures, enhancing its utility in various fields.

In many practical applications, the stochastic block structure may be more relevant for correlation matrices than covariance matrices. For example, principle component analysis and copula models rely on correlation matrices. In portfolio selection, some assets may exhibit large variances, which can dominate clustering results and obscure important co-movements. When clustering is performed based on the covariance matrix, variables with large variances often form distinct blocks, pushing variables with smaller variances into a single block. This sensitivity to variance can lead to suboptimal clustering that fails to capture meaningful patterns of co-fluctuation.

To more accurately capture the block structure in a correlation matrix, we recommend standardizing the data to have zero mean and unit variance before applying the stochastic block covariance model. However, standardizing data is not always feasible, as correlation matrices may be an intermediate product rather than the target estimator. Furthermore, standardization can ignore the uncertainty quantification of variance estimation by treating variance as fixed. A more direct approach is to consider a semi-spectral decomposition of block correlation matrices. However, this introduces computational challenges due to the inherent restriction that correlations $r_{ij} \in [-1,1]$, imposing constraints on the parameters $(A,\mathbf{\lambda})$. These constraints make it difficult to apply standard techniques, as a conjugate prior is not readily available, complicating both the estimation of correlations and the block assignment process. A better solution might involve introducing scaling parameters to account for heteroskedasticity. We can express the covariance matrix as $\Gamma = V^{1/2}R V^{1/2}$, where $V$ is a diagonal matrix representing individual variances, and $\Sigma$ is a block covariance matrix. 


A substantial challenge to our method is the computational cost which scales poorly with large datasets. Our current implementation cannot handle datasets with $p > 200$, as often encountered in proteomics, gene expression, and finance. The bottleneck lies in estimating the block assignments under partition priors, where even with an effective merge-split sampler \citep{dahl2022sequentially}, we encounter slow mixing and high computational costs. It would be worthwhile to innovate more efficient computing strategies which can overcome the incremental nature of present Markov chain samplers in exploring the space of partitions.

\section{Acknowledgement}

We thank Marlene Cohen, Douglas Ruff, Silvia Matesanz, Micaela Carvajal, Jonathan Azose, and Jason Xu for sharing their data with us. We also thank Jaeho Kim and Drew Creal for providing their code and detailed numerical experiment settings. We benefitted from helpful discussion with Ed Tam and Johan Vdmolen. The work was partially supported by NIH awards R01 DC013096 and R01 DC016363.

\newpage
\bibliographystyle{chicago}

\newpage
\appendix

\section*{Supplementary Materials}
\section{Obtain rotation matrix Q}
\label{append:a}
\begin{equation}
Q^{(p_k)}=\Tilde{Q}^{(p_k)}\begin{pmatrix}
1/\sqrt{p_k} &  &  & \\
 & 1/\sqrt{1+1^2} & & \\
 & & \ddots & \\
 & & & 1/\sqrt{i+i^2}\\
 & & & & \ddots & \\
 & & & & & 1/\sqrt{(p_k-1)+(p_k-1)^2}  \\
\end{pmatrix}
\label{equ:Q1}
\end{equation}

\begin{equation}
\Tilde{Q}^{(p_k)}=(\Tilde{Q}_1^{(p_k)}~\Tilde{Q}_{-1}^{(p_k)})=\begin{pmatrix}
1&-1&1 &1&\cdots&1\\
1&0 &-2&1&\cdots&1\\
1&0 &0 &-3& \ddots &1 \\
1&0 &0 &0 & \ddots &1 \\
1&0 &0 &0 & \cdots &-(p_k-1) \\
1&1 &1 &1 & \cdots &1\\
\end{pmatrix}
\label{equ:Q2}
\end{equation}

\section{Proof of Theorem~\ref{theorem1}}
\label{append:b}
From Theorem 3 in \cite{archakov2024canonical}, we have maximum likelihood estimator for $A$ and $\lambda_k~(k \in \{1,\dots,K\})$ as 

\begin{equation}
\tag{9}
    \hat{A}=\frac{\sum_{i=1}^n\bm{\eta}_{i(0)}\bm{\eta}_{i(0)}'}{n}, \quad \hat{\lambda}_k=\frac{\sum_{i=1}^n\bm{\eta}_{i(k)}'\bm{\eta}_{i(k)}}{n(p_k-1)}
\end{equation}

Since $\Sigma$ is a rotation of $D$, the MLE of $\Sigma$ is obtained by plugging $\hat{A}$ and $\hat{\lambda}_k$ into equation (\ref{equ:cr}).
\begin{equation}
    \hat{\Sigma}=Q\hat{D}Q'=Q\begin{pmatrix}
\hat{A} & 0 & \dots & 0\\
0 & \hat{\lambda}_1 I_{p_1-1} & \ddots& \vdots\\
\vdots &\ddots & \ddots & 0 \\
0 & \dots & 0 & \hat{\lambda}_K I_{p_K-1}
\end{pmatrix}Q'
\end{equation}

Recall the equations (\ref{equ:al}), we can represent entries of $\Sigma$ using entries of $A$ and $\lambda_k$ as follows:

\begin{equation}
\label{equ:sigma_al}
    \sigma_k^2=\frac{a_{kk}+(p_k-1)\lambda_k}{p_k},~ \sigma_{kk}=\frac{a_{kk}-\lambda_k}{p_k},~\sigma_{kl}=\frac{a_{kl}}{\sqrt{p_kp_l}}
\end{equation}

Plugging equation (\ref{equ:mle}) to equation (\ref{equ:sigma_al}), we have 

\begin{align}
    \hat{\sigma}_k^2=&\frac{1}{np_k}\sum_{i=1}^n(\eta_{i(0),k}^2+\bm{\eta}_{i(k)}'\bm{\eta}_{i(k)})\label{equ:sigma_k}\\
    \hat{\sigma}_{kk}=&\frac{1}{np_k}\sum_{i=1}^n(\bm{\eta}_{i(0),k}^2-\frac{1}{p_k-1}\bm{\eta}_{i(k)}'\bm{\eta}_{i(k)})\\
    \hat{\sigma}_{kl}=&\frac{1}{n\sqrt{p_kp_l}}\sum_{i=1}^n\bm{\eta}_{i(0),k}'\bm{\eta}_{i(0),l}\label{equ:sigma_kl}    
\end{align}

Let $\Tilde{\bm{y}}_i$ denotes $i^{th}$ observation, which is a sorted $p\times 1$ vector. We have $\Tilde{\bm{y}}_i=(\Tilde{\bm{y}}_i^{(p_1)'}, \dots,$ $\Tilde{\bm{y}}_i^{(p_K)'})'$, where $\Tilde{\bm{y}}_i^{(p_k)}$ is a $p_k \times 1$ data vector which belongs to block $k$. Consequently, We can partition $Q'\Tilde{\bm{y}}_i$ as follows:
\begin{equation}
    Q'\Tilde{\bm{y}}_i=(Q_1^{(p_1)'}\Tilde{\bm{y}}_i^{(p_1)},\dots,Q_1^{(p_K)'}\Tilde{\bm{y}}_i^{(p_K)},Q_{-1}^{(p_1)'}\Tilde{\bm{y}}_i^{(p_1)},\dots,Q_{-1}^{(p_K)'}\Tilde{\bm{y}}_i^{(p_K)})
\end{equation}

Notice $Q'\Tilde{\bm{y}}_i=\bm{\eta}_i'=(\bm{\eta}_{i(0)}',\bm{\eta}_{i(1)}',\dots,\bm{\eta}_{i(k)}',\dots,\bm{\eta}_{i(K)}')$, $\bm{\eta}_{i(0)} \in \mathbb{R}^{K\times1}$, and $\bm{\eta}_{i(k)} \in \mathbb{R}^{(p_k-1)\times1}$. Then we have 
\begin{equation}
    \begin{split}
        \bm{\eta}_{i(0)}'=(Q_1^{(p_1)'}\Tilde{\bm{y}}_i^{(p_1)},\dots,Q_1^{(p_K)'}\Tilde{\bm{y}}_i^{(p_K)})\\
    \bm{\eta}_{i(k)}'=Q_{-1}^{(p_k)'}\Tilde{\bm{y}}_i^{(p_k)}, ~ k \in \{1,\dots,K\}\label{equ:eta_yk}        
    \end{split}
\end{equation}

Plug equations (\ref{equ:eta_yk}) into equations (\ref{equ:sigma_k})-(\ref{equ:sigma_kl}), we can directly match the MLE of $\Sigma$ back to MLE of pairwise covariance. 

\begin{align*}
\hat{\sigma}_k^2&=\frac{1}{np_k}\sum_{i=1}^n\bm{y}_{i}^{(k)'}Q^{(p_k)}Q^{(p_k)'}\bm{y}_{i}^{(k)}=\frac{1}{np_k}\sum_{i=1}^n\bm{y}_{i}^{(k)'}\bm{y}_{i}^{(k)}=\frac{1}{np_k}\sum_{i=1}^n\sum_{j\in \mathcal{B}_k}y_{ij}^2\\
\hat{\sigma}_{kk}&=\frac{1}{np_k}\sum_{i=1}^n(\bm{y}_{i}^{(k)'}Q_1^{(p_k)}Q_1^{(p_k)'}\bm{y}_{i}^{(k)}-\frac{1}{(p_k-1)}\bm{y}_{i}^{(k)'}Q_{-1}^{(p_k)}Q_{-1}^{(p_k)'}\bm{y}_{i}^{(k)})\\
&=\frac{1}{np_k(p_k-1)}\sum_{i=1}^n\bm{y}_{i}^{(k)'}(\frac{1}{p_k}\bm{1}\bm{1}'-\frac{1}{p_k-1}(I-\frac{1}{p_k}\bm{1}\bm{1}'))\bm{y}_{i}^{(k)}\\
&=\frac{1}{np_k(p_k-1)}\sum_{i=1}^n\bm{y}_{i}^{(k)'}(\bm{1}\bm{1}'-I)\bm{y}_{i}^{(k)}=\frac{1}{np_k(p_k-1)}\sum_{i=1}^n\sum_{j,j'\in \mathcal{B}_k,j\neq j'}y_{ij}y_{ij'}\\
\hat{\sigma}_{kl}&=\frac{1}{np_kp_l}\sum_{i=1}^n\bm{y}_{i}^{(k)'}Q_1^{(p_k)}Q_1^{(p_l)'}\bm{y}_{i}^{(l)}=\frac{1}{np_kp_l}\sum_{i=1}^n\bm{y}_{i}^{(k)'}\bm{1}\bm{1}'\bm{y}_{i}^{(l)}=\frac{1}{np_kp_l}\sum_{i=1}^n\sum_{j\in \mathcal{B}_k,j'\in \mathcal{B}_l}y_{ij}y_{ij'}
\end{align*}
\newpage


        
        
        



\section{Expanded MCMC for the hierarchical method}
\label{sampler:expand-mcmc}

Let $\boldsymbol{\theta}$ denote the hyperparameters $(\nu_0,s_0,\delta_1,\delta_2,\delta_3)$. We adopt a block-update strategy, dividing the sampling process into three major steps:

Start with initial values $\mathcal{B}^{(0)}$ and $\theta^{(0)}$, for each iteration $t \in \{1,\dots,T\}$:

\begin{enumerate}
    \item Given $(\boldsymbol{\theta}^{(t-1)})$, update $\mathcal{B}^{(t)}$ using the Sequentially Allocated Merge-Split Sampler (SAMS) introduced by \cite{dahl2022sequentially}.
    \item Given $(\mathcal{B}^{(t)},A^{(t)},\boldsymbol{\lambda}^{(t)})$, update $\boldsymbol{\theta}^{(t)}$ using the Adaptive Metropolis algorithm with global adaptive scaling according to Algorithm 4 from \cite{andrieu2008tutorial}.
\end{enumerate}
Once samples of $(\sc B, \bm\theta)$ are drawn from the posterior, those of $\Gamma$ could be drawn easily by using the conjugate form described in \eqref{equ:A}.

\newpage
\section{Extra figures for Numerical Experiments}

\begin{figure}[ht]
\begin{center}
\includegraphics[width=0.9\linewidth]{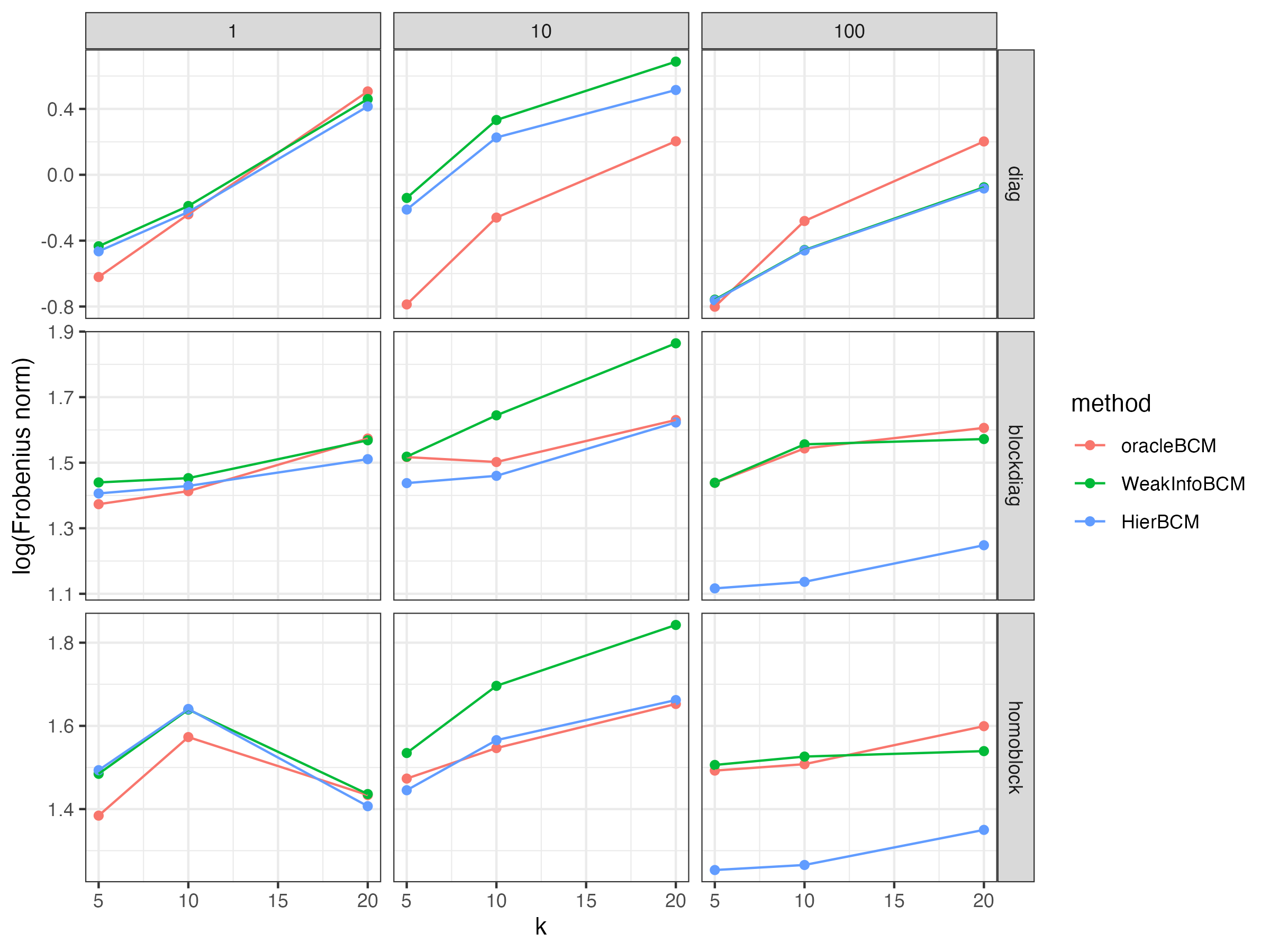}    
\end{center}
\caption{Comparison of estimation accuracy among different priors under well-specified cases ($p=50$).}
\label{fig:well.n50.fn}
\end{figure}

\begin{figure}[ht]
\begin{center}
\includegraphics[width=0.9\linewidth]{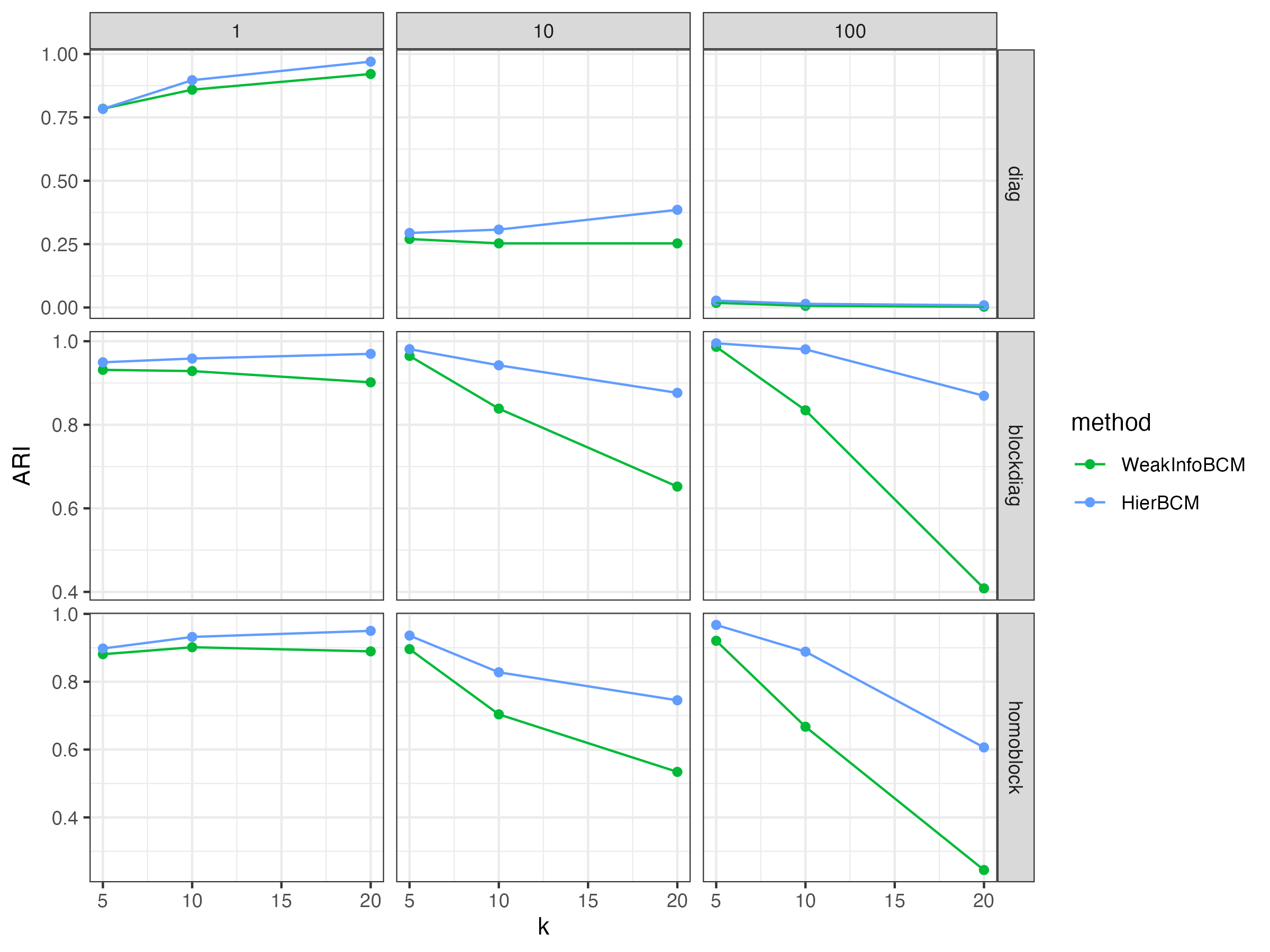}
\end{center}
\caption{Comparison of clustering performance among different priors under well-specified cases ($n=50$).}
\label{fig:well.n50.ari}
\end{figure}

\begin{figure}[ht]
\begin{center}
\includegraphics[width=0.9\linewidth]{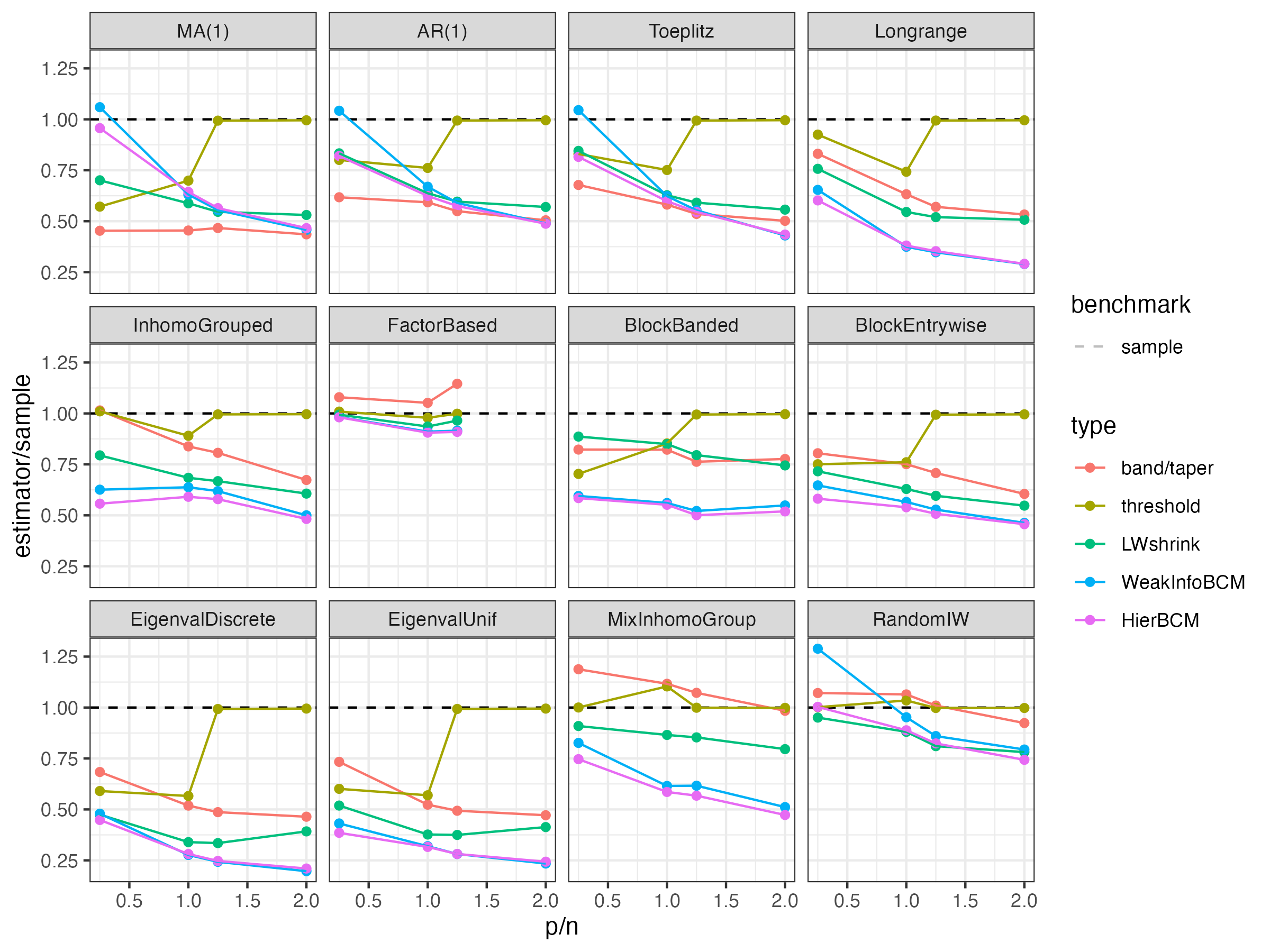}
\end{center}
\captionsetup{justification=raggedright,singlelinecheck=false}
\caption{Estimation accuracy of different estimators across various covariances}
\label{fig:rfn.p20}
\end{figure}

\clearpage
\pagebreak

\section{Extra tables for Numerical Experiments}

Table \ref{tab:well-summary} present a summary statistics (posterior mean of Frobenius distance, number of blocks, ARI, R2) calculated from posterior samples. Notice R2 is a metric measuring clustering accuracy proposed by \cite{creal2024bayesian}, as defined below:

\begin{equation}
    R_2 = E(L(\hat{\Pi},\Pi_0)) = E(-\frac{1}{p^2}\sum_{i=1}^p\sum_{i'=1}^p \pi_{0,ii'}\log(\hat{\pi}_{ii'}))
\end{equation}

\noindent where $\hat{\Pi},\Pi_0$ represent estimator and the true probability matrix respectively. Each entry $\hat{\pi}_{ii'},\pi_{0,ii'}$ of the matrix $\hat{\Pi},\Pi_0$ quantify the probability individual $i$ and $i'$ belong to the same cluster. We consider ARI instead of R2 since R2 is sensitive to number of clusters and size of clusters. We can observes some cases that better estimation on number of blocks does not appear to be better in the performance measured by R2. R2 is not consistent with the $\hat{K}$.

\relax

\begin{table}[ht]
\centering
    \caption{Comparison of priors in well-specified scenarios}
    \resizebox{\textwidth}{!}{
    \begin{tabular}{llllllllllllll}
        \toprule
        \textbf{Dataset} & \textbf{n} & \textbf{Prec} & \textbf{k} & \textbf{Sample} & \textbf{Oracle} & \textbf{Weak} & \textbf{BCM} & \textbf{K\_Weak} & \textbf{K\_BCM} & \textbf{ARI\_Weak} & \textbf{ARI\_BCM} & \textbf{R2\_Weak} & \textbf{R2\_BCM} \\
        \midrule
        \multirow{18}{*}{diagonal} & \multirow{9}{*}{25} & \multirow{3}{*}{1} & 5 & 4.725 & 0.751 & 1.031 & 1.01 & 3.735 & 4.198 & 0.541 & 0.536 & 0.257 & 0.101 \\
         & & & 10 & 4.288 & 1.282 & 1.563 & 1.486 & 6.31 & 7.298 & 0.557 & 0.58 & 0.174 & 0.099 \\
         & & & 20 & 3.952 & 1.713 & 2.057 & 1.929 & 10.648 & 12.649 & 0.56 & 0.623 & 0.22 & 0.065 \\
        \cmidrule(lr){3-14}
         & & \multirow{3}{*}{10} & 5 & 4.895 & 0.652 & 1.116 & 1.09 & 2.153 & 2.835 & 0.145 & 0.153 & 0.145 & 0.145 \\
         & & & 10 & 5.132 & 1.091 & 1.623 & 1.546 & 2.499 & 3.493 & 0.113 & 0.131 & 0.12 & 0.099 \\
         & & & 20 & 5.233 & 1.875 & 2.576 & 2.405 & 3.548 & 5.142 & 0.103 & 0.124 & 0.139 & 0.091 \\
        \cmidrule(lr){3-14}
         & & \multirow{3}{*}{100} & 5 & 5.218 & 0.614 & 0.511 & 0.539 & 1.212 & 1.823 & 0.004 & 0.008 & 0.017 & 0.066 \\
         & & & 10 & 5.101 & 1.037 & 0.654 & 0.674 & 1.189 & 1.762 & 0.001 & 0.003 & 0.014 & 0.044 \\
         & & & 20 & 5.164 & 1.734 & 0.948 & 0.966 & 1.237 & 1.906 & 0.002 & 0.004 & 0.008 & 0.028 \\
        \cmidrule(lr){2-14}
         & \multirow{9}{*}{50} & \multirow{3}{*}{1} & 5 & 3.568 & 0.537 & 0.648 & 0.628 & 4.274 & 4.653 & 0.784 & 0.783 & 0.066 & 0.059 \\
         & & & 10 & 3.541 & 0.786 & 0.827 & 0.798 & 8.201 & 8.951 & 0.859 & 0.897 & 0.019 & 0.016 \\
         & & & 20 & 5.15 & 1.659 & 1.584 & 1.514 & 14.939 & 16.052 & 0.921 & 0.97 & 0.002 & 0.001 \\
        \cmidrule(lr){3-14}
         & & \multirow{3}{*}{10} & 5 & 3.304 & 0.455 & 0.869 & 0.81 & 2.343 & 3.241 & 0.27 & 0.294 & 0.148 & 0.151 \\
         & & & 10 & 3.592 & 0.771 & 1.395 & 1.254 & 3.306 & 5.023 & 0.253 & 0.307 & 0.152 & 0.123 \\
         & & & 20 & 3.61 & 1.226 & 1.988 & 1.674 & 4.849 & 8.952 & 0.253 & 0.385 & 0.147 & 0.106 \\
        \cmidrule(lr){3-14}
         & & \multirow{3}{*}{100} & 5 & 3.64 & 0.449 & 0.468 & 0.466 & 1.195 & 1.715 & 0.018 & 0.027 & 0.056 & 0.06 \\
         & & & 10 & 3.624 & 0.755 & 0.633 & 0.631 & 1.207 & 1.831 & 0.006 & 0.014 & 0.026 & 0.051 \\
         & & & 20 & 3.612 & 1.224 & 0.927 & 0.92 & 1.247 & 2.043 & 0.003 & 0.009 & 0.016 & 0.036 \\
        \midrule
        \multirow{18}{*}{blockdiag} & \multirow{9}{*}{25} & \multirow{3}{*}{1} & 5 & 8.328 & 4.193 & 4.911 & 4.852 & 4.379 & 4.715 & 0.84 & 0.864 & 0.028 & 0.023 \\
         & & & 10 & 9.434 & 6.002 & 6.281 & 5.973 & 6.757 & 7.741 & 0.741 & 0.801 & 0.096 & 0.034 \\
         & & & 20 & 9.51 & 6.543 & 6.922 & 6.504 & 10.801 & 12.981 & 0.658 & 0.75 & 0.104 & 0.026 \\
        \cmidrule(lr){3-14}
         & & \multirow{3}{*}{10} & 5 & 10.453 & 6.087 & 6.657 & 5.986 & 3.978 & 4.74 & 0.839 & 0.904 & 0.016 & 0.027 \\
         & & & 10 & 10.785 & 6.672 & 7.513 & 6.47 & 4.926 & 6.984 & 0.661 & 0.813 & 0.058 & 0.028 \\
         & & & 20 & 10.216 & 6.772 & 8.219 & 7.16 & 5.182 & 8.515 & 0.357 & 0.563 & 0.107 & 0.038 \\
        \cmidrule(lr){3-14}
         & & \multirow{3}{*}{100} & 5 & 10.314 & 5.764 & 6.032 & 4.187 & 4.093 & 4.979 & 0.887 & 0.958 & 0.043 & 0.012 \\
         & & & 10 & 10.686 & 6.574 & 6.61 & 4.793 & 3.846 & 6.878 & 0.562 & 0.847 & 0.024 & 0.022 \\
         & & & 20 & 10.319 & 6.854 & 6.102 & 5.212 & 3.296 & 7.029 & 0.207 & 0.543 & 0.024 & 0.023 \\
        \cmidrule(lr){2-14}
         & \multirow{9}{*}{50} & \multirow{3}{*}{1} & 5 & 9.437 & 3.949 & 4.22 & 4.08 & 4.624 & 4.905 & 0.931 & 0.95 & 0.008 & 0.012 \\
         & & & 10 & 6.854 & 4.109 & 4.275 & 4.175 & 8.451 & 9.1 & 0.929 & 0.959 & 0.006 & 0.006 \\
         & & & 20 & 6.548 & 4.825 & 4.799 & 4.53 & 14.433 & 16.211 & 0.902 & 0.97 & 0.006 & 0.004 \\
        \cmidrule(lr){3-14}
         & & \multirow{3}{*}{10} & 5 & 7.587 & 4.559 & 4.565 & 4.211 & 4.655 & 4.981 & 0.965 & 0.981 & 0.002 & 0.004 \\
         & & & 10 & 7.317 & 4.491 & 5.178 & 4.306 & 6.215 & 8.413 & 0.839 & 0.942 & 0.004 & 0.007 \\
         & & & 20 & 7.61 & 5.106 & 6.451 & 5.067 & 7.74 & 13.082 & 0.652 & 0.876 & 0.03 & 0.009 \\
        \cmidrule(lr){3-14}
         & & \multirow{3}{*}{100} & 5 & 7.402 & 4.215 & 4.215 & 3.054 & 4.782 & 5.051 & 0.987 & 0.995 & 0.002 & 0.003 \\
         & & & 10 & 7.494 & 4.681 & 4.74 & 3.116 & 5.322 & 8.431 & 0.835 & 0.981 & 0.001 & 0.001 \\
         & & & 20 & 7.326 & 4.984 & 4.817 & 3.484 & 4.826 & 11.138 & 0.408 & 0.869 & 0.014 & 0.003 \\
        \midrule
        \multirow{18}{*}{centerblock} & \multirow{9}{*}{25} & \multirow{3}{*}{1} & 5 & 41.761 & 8.625 & 9.467 & 9.149 & 4.207 & 4.594 & 0.789 & 0.809 & 0.069 & 0.045 \\
         & & & 10 & 10.466 & 6.851 & 7.084 & 6.752 & 6.666 & 7.555 & 0.705 & 0.755 & 0.075 & 0.05 \\
         & & & 20 & 9.423 & 5.63 & 6.382 & 6.139 & 10.08 & 11.855 & 0.613 & 0.683 & 0.096 & 0.037 \\
        \cmidrule(lr){3-14}
         & & \multirow{3}{*}{10} & 5 & 10.214 & 5.548 & 5.996 & 5.442 & 3.488 & 4.365 & 0.685 & 0.765 & 0.046 & 0.056 \\
         & & & 10 & 10.772 & 6.815 & 7.339 & 6.591 & 3.942 & 5.498 & 0.481 & 0.606 & 0.061 & 0.052 \\
         & & & 20 & 10.317 & 7.043 & 8.296 & 7.62 & 4.58 & 6.655 & 0.288 & 0.416 & 0.103 & 0.059 \\
        \cmidrule(lr){3-14}
         & & \multirow{3}{*}{100} & 5 & 10.685 & 6.132 & 6.301 & 4.91 & 3.144 & 4.384 & 0.638 & 0.8 & 0.04 & 0.045 \\
         & & & 10 & 10.575 & 6.454 & 6.256 & 5.306 & 2.669 & 4.773 & 0.317 & 0.551 & 0.036 & 0.057 \\
         & & & 20 & 10.418 & 6.956 & 5.817 & 5.307 & 2.24 & 4.225 & 0.095 & 0.246 & 0.025 & 0.038 \\
        \cmidrule(lr){2-14}
         & \multirow{9}{*}{50} & \multirow{3}{*}{1} & 5 & 7.513 & 3.992 & 4.414 & 4.453 & 4.509 & 4.813 & 0.881 & 0.898 & 0.023 & 0.028 \\
         & & & 10 & 7.064 & 4.821 & 5.151 & 5.158 & 8.255 & 8.949 & 0.902 & 0.932 & 0.01 & 0.009 \\
         & & & 20 & 6.14 & 4.194 & 4.204 & 4.084 & 14.29 & 15.864 & 0.89 & 0.95 & 0.018 & 0.012 \\
        \cmidrule(lr){3-14}
         & & \multirow{3}{*}{10} & 5 & 7.563 & 4.364 & 4.64 & 4.243 & 4.313 & 4.842 & 0.896 & 0.936 & 0.006 & 0.015 \\
         & & & 10 & 7.553 & 4.695 & 5.453 & 4.786 & 5.205 & 7.293 & 0.704 & 0.828 & 0.04 & 0.05 \\
         & & & 20 & 7.673 & 5.221 & 6.313 & 5.27 & 6.851 & 11.927 & 0.534 & 0.745 & 0.054 & 0.027 \\
        \cmidrule(lr){3-14}
         & & \multirow{3}{*}{100} & 5 & 7.589 & 4.448 & 4.509 & 3.502 & 4.201 & 4.967 & 0.921 & 0.967 & 0.008 & 0.008 \\
         & & & 10 & 7.427 & 4.517 & 4.601 & 3.545 & 3.986 & 6.814 & 0.667 & 0.889 & 0.007 & 0.01 \\
         & & & 20 & 7.329 & 4.95 & 4.661 & 3.857 & 3.49 & 7.81 & 0.245 & 0.606 & 0.024 & 0.015 \\
        \bottomrule
    \end{tabular}        
    }
    \label{tab:well-summary}
\end{table}

\clearpage
\pagebreak

\section{More details for Applications}

\subsection{Preprocessing procedure of neuroscience application}
\label{append:neuro}

We collect spike counts within a window spanning from 30 ms to 230 ms. Our analysis is confined to the experimental setting, where a certain percentage of neurons exhibit code juggling patterns when encoding dual stimuli. Our attention is specifically directed to a subset of neurons demonstrating code juggling, as identified by the Spike Count Analysis for MultiPlexing Inference (SCAMPI) model \citep{chen2024spike}. The raw data consists of spike count data, which does not inherently reveal the preference for selecting between two stimuli. Therefore, we employ a Poisson mixture model to represent a neuron's spike count, incorporating a mixing proportion $\alpha$ to quantify the selection preference between stimuli $\A$ and $\B$:

\begin{equation}
    Y^{\AB}|\mu^\A,\mu^\B \sim \int_0^1 \poi(\alpha\mu^\A+(1-\alpha)\mu^\B)f(\alpha) d\alpha 
    \label{equ:int}
\end{equation}

The model shown in equation (\ref{equ:int}) is exactly the model under Slow Juggling hypothesis in the SCAMPI model. We applied SCAMPI model and obtain an estimation of mixing density $f(\alpha)$, and estimation (posterior mode) of parameters $\mu^\A$ and $\mu^\A$. Then we derive the posterior of $\alpha$ based on Bayes' rule, utilizing the posterior mean as an estimator of $\alpha_i$ for each neuron at trial $i$. This process yields a score matrix representing the selection preference of 74 neurons across 20 trials:

\begin{equation}
    p(\alpha|Y^\AB_i) = \frac{\poi(Y_i^\AB|\alpha\hat{\mu}^\A+(1-\alpha)\hat{\mu}^\B)f(\alpha)}{\int_0^1\poi(Y_i^\AB|\alpha\hat{\mu}^\A+(1-\alpha)\hat{\mu}^\B)f(\alpha)d\alpha}
\end{equation}

\subsection{Extra figures from applications}
\label{append:app}

\begin{figure}[ht]
\begin{center}
\includegraphics[width=0.8\linewidth]{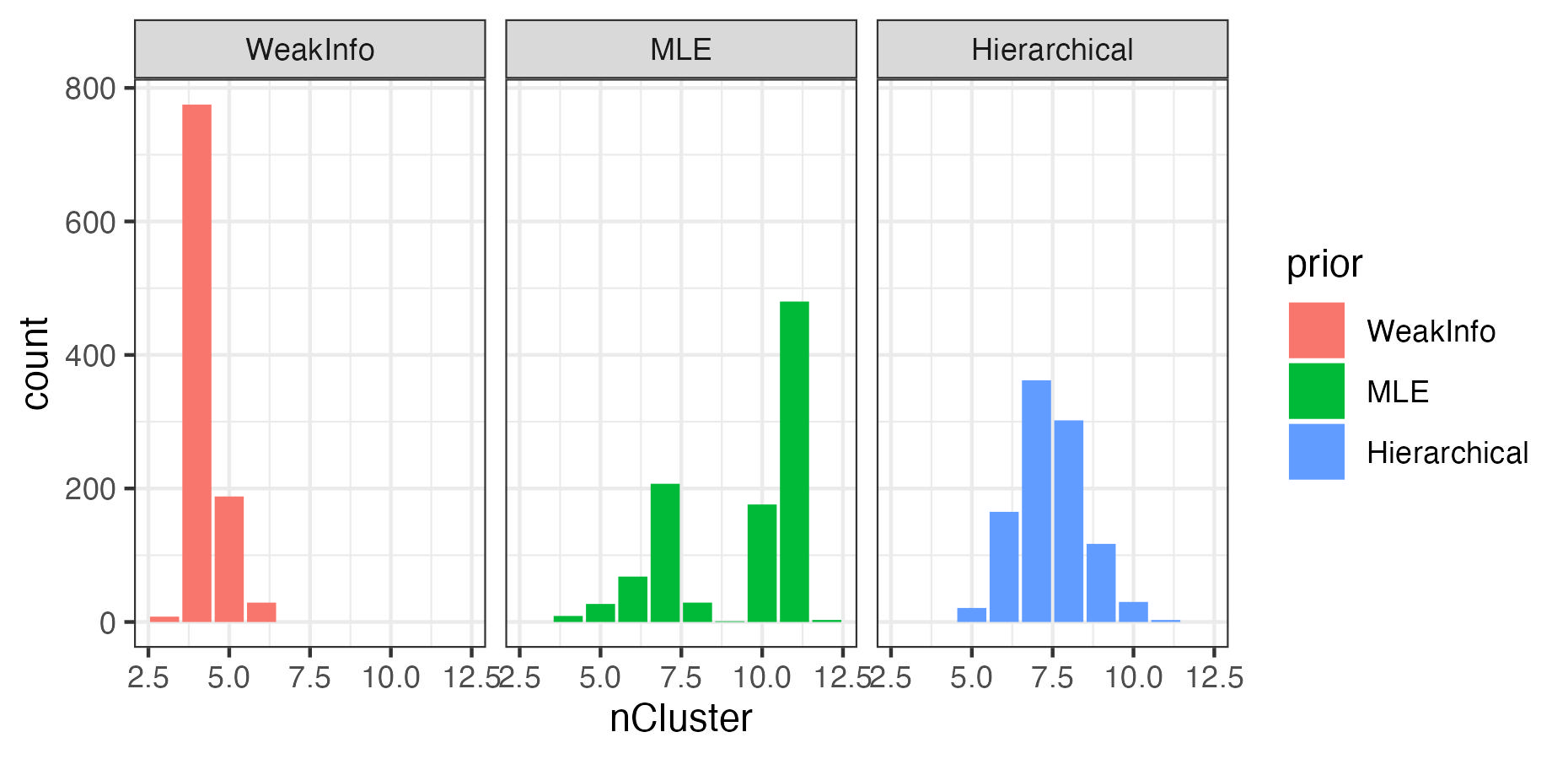}
\end{center}
\caption{Posterior distributions of the number of blocks across different priors in the neural population coordination application.}
\label{fig:neuron_k}
\end{figure}

\clearpage

\begin{figure}[!ht]
\begin{center}
\includegraphics[width=0.8\linewidth]{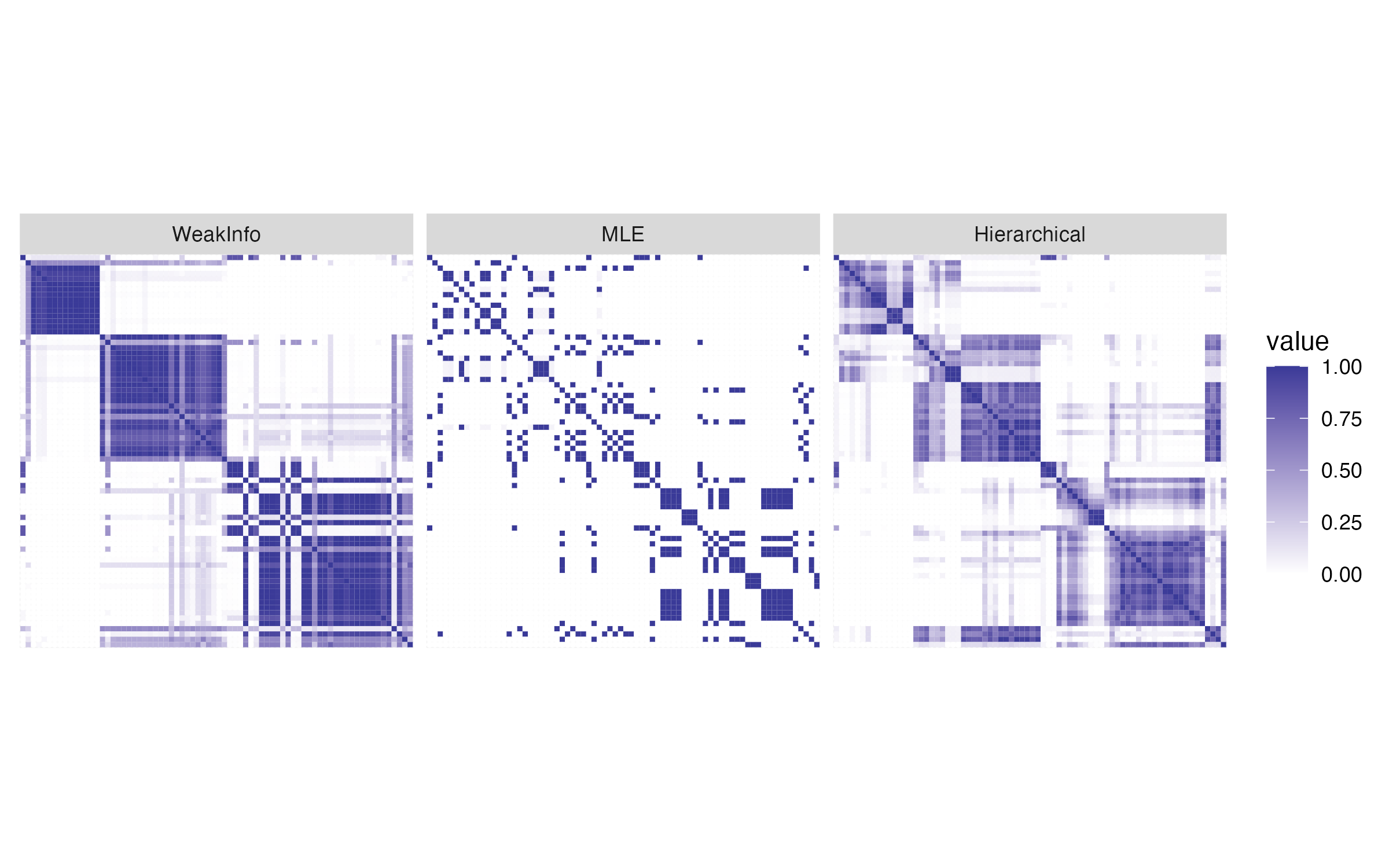}    
\end{center}
\caption[Posterior similarity matrices under different priors.]{Posterior similarity matrices under different priors. The weakly informative prior oversimplifies the block structure, failing to capture finer details. The MLE-equivalent prior exhibits slow MCMC mixing or may become trapped in a local mode, affecting accurate block identification.}
\label{fig:neuron_psm}
\end{figure}

\begin{figure}[!ht]
\begin{center}
\includegraphics[width=0.8\linewidth]{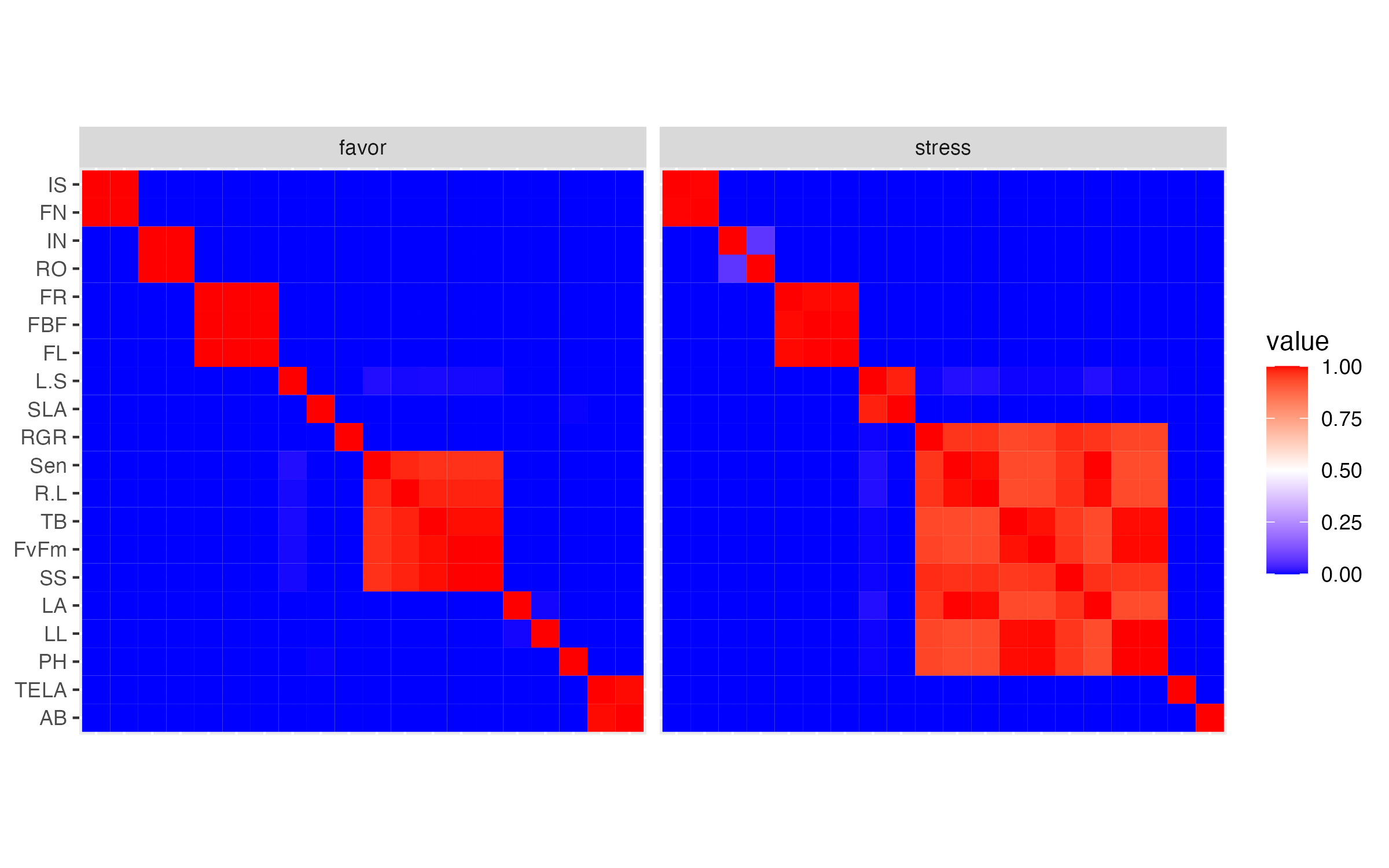}
\end{center}
\caption[Posterior similarity matrices of traits under favorable and stressful environments.]{Posterior similarity matrices of traits under favorable and stressful environments. The number of distinct clusters decreases under drought, indicating stronger phenotypic integration.}
\label{fig:plant_psm}
\end{figure}

\end{document}